\documentclass[12pt]{article}

\input tcilatex

\begin{document}

\title{Nonlinear Hodge maps}
\author{Thomas H. Otway\thanks{%
email: otway@ymail.yu.edu. This is a slightly modified version of an article
which appeared in the \textit{Journal of Mathematical Physics,} Volume 41,
5745 (2000) and may be found at http://link.aip.org/link/?jmp/41/5745.
\copyright 2000 American Institute of Physics. This article may be
downloaded for personal use only. Any other use requires prior permission of
the American Institute of Physics.} \\
\\
\textit{Department of Mathematics and }\\
\textit{Department of Physics, Yeshiva University,}\\
\textit{500 W 185th Street, New York, New York 10033}}
\date{}
\maketitle

\begin{abstract}
We consider maps between Riemannian manifolds in which the map is a
stationary point of the nonlinear Hodge energy. \ The variational equations
of this functional form a quasilinear, nondiagonal, nonuniformly elliptic
system which models certain kinds of compressible flow. \ Conditions are
found under which singular sets of prescribed dimension cannot occur.\ \
Various degrees of smoothness are proven for the sonic limit,
high-dimensional flow, and flow having nonzero vorticity. \ The gradient
flow of solutions is estimated. \ Implications for other quasilinear field
theories are suggested.
\end{abstract}

\newpage

\section{Introduction: nonlinear Hodge theory}

The original motivation for nonlinear Hodge theory was a conjecture of Bers
on the existence of subsonic compressible flow having prescribed periods on
a Riemannian manifold. \ In establishing Bers' conjecture for a stationary,
irrotational, polytropic flow,$^{1}$ L. M. and R. J. Sibner were led to more
general questions about the properties of differential forms on Riemannian
manifolds.$^{2-4}$ \ In Ref. 2 they introduced a variational principle for
the generalized energy functional 
\begin{equation}
E=\frac{1}{2}\int_{M}\int_{0}^{Q}\rho (s)ds\,dM.
\end{equation}
(See also Ref. 5, p. 221.) \ Here $M$ is a smooth $n$-dimensional Riemannian
manifold. \ Denote by $\omega $ a smooth section of the $p^{th}$ exterior
power of the cotangent bundle on $M;$ then $Q(\omega )$ is the square of the
pointwise norm of $\omega $ on $TM$. The $C^{1}$ function $\rho :\mathbf{R}%
\rightarrow \mathbf{R}^{+}$ is assumed to satisfy the conditions 
\begin{equation}
0<\frac{d}{dQ}\left( Q\rho ^{2}(Q)\right) <\infty
\end{equation}
for $Q<Q_{crit}$, and 
\begin{equation}
\lim_{Q\rightarrow Q_{crit}}\frac{d}{dQ}\left( Q\rho ^{2}(Q)\right) =0.
\end{equation}
If $p=1$ and $\omega $ is the 1-form canonically associated to the velocity
field of an adiabatic, isentropic compressible flow on $M$, then $\rho $ is
given explicitly by the formula 
\[
\rho (Q)=\left( 1-\frac{\gamma _{a}-1}{2}Q\right) ^{1/(\gamma _{a}-1)}, 
\]
where $\gamma _{a}>1$ is the adiabatic constant of the medium, and $%
Q_{crit}=2/(\gamma _{a}+1)$ is the square of the sonic flow velocity.

The variational equations of $E$ are the \textit{nonlinear Hodge equations} 
\begin{equation}
\delta \left( \rho (Q)\omega \right) =0.
\end{equation}
If the flow is irrotational then we have an additional equation 
\begin{equation}
d\omega =0.
\end{equation}
Here $d:\Lambda ^{p}\rightarrow \Lambda ^{p+1}$ is the (flat) exterior
derivative on \textit{p}-forms, with adjoint operator $\delta :\Lambda
^{p}\rightarrow \Lambda ^{p-1}.$

Applying the converse of the Poincar\'{e} Lemma to eq. (5), we find that,
locally, there is a $p-1$-form $u$ such that $du=\omega .$

It has been observed$^{2}$ that if $\{x_{1},...,x_{n+1}\}$ are coordinates
in $\mathbf{R}^{n+1},$ $u$ is a mapping of $M$ into $\mathbf{R}^{n+1}$ such
that $x_{n+1}=$ $u(x_{_{1}},...,x_{n}),$ $p=1,$ and 
\begin{equation}
\rho (Q)=(1+Q)^{-1/2},
\end{equation}
then eq. (4) can be interpreted as the equation for a family of
codimension-1 minimal hypersurfaces having gradients $\nabla u.$ \ The
critical value of $Q$ is $Q=\infty .$ \ Of course $E$ does not yield the
area functional but rather an indefinite functional 
\begin{equation}
\int_{M}\left( \sqrt{1+Q}-1\right) dM
\end{equation}
which differs from the area functional by an integration constant.

If $p=2,\,n=4,$ and $\omega $ denotes an electromagnetic field having
electromagnetic potential $u,$ then for $\rho =1$ eqs. (4), (5) reduce to
Maxwell's equations on $M$. \ \ If, however, we replace the standard model
of electromagnetism with the Born-Infeld model, then we have (taking the
energy to be positive-definite) 
\begin{equation}
E_{Born-Infeld}=b^{2}\int_{M}\left( \sqrt{1+\frac{1}{2b^{2}}Q}-1\right) dM,
\end{equation}
where $b^{2}=mc^{2}.$ \ The nonzero integration constant observed in (7)
arises in (8) from independent physical arguments [\textit{c.f.} \ eqs.
(1.1) and (1.4) of Ref. 6]. \ Normalizing so that $b^{2}=1/2$, the energy
functional (8) becomes identical to the functional (7), and we can choose $%
\rho $ as in eq. (6) in order to write the variational equations of this
common energy functional in the form of eqs. (4). \ Thus the Born-Infeld
model fits naturally into nonlinear Hodge theory as an application for
2-forms. \ (The equations for a nonparametric codimension-1 minimal surface
also have a place in the original gas dynamics context of nonlinear Hodge
theory, as the Chaplygin approximation of a compressible flow.$^{7}$)

Because the bundle $T^{\ast }M$ is flat, any connection defined on it will
have trivial Lie bracket. \ For this reason, in comparison to the examples
that follow, we call the foregoing cases \textit{abelian.} \ In particular,
in the example of electromagnetism the vector potential $u$ is identified
with a connection 1-form on a bundle over $M$ having abelian structure group 
$U(1) $.

Suppose we replace the energy functionals (1) and (7) by the functional 
\[
E_{h}=\frac{1}{2}\int_{M}\int_{-h}^{Q}\rho (s)ds\,dM, 
\]
where 
\[
\rho (s)=(h+s)^{-\alpha }. 
\]
Here $h$ and $\alpha $ are nonnegative parameters. \ Let $X$ be a vector
bundle over $M$ having compact structure group $G$. \ Define $Q=\left\langle
F,F\right\rangle ,$ where $\left\langle \;,\;\right\rangle $ is an inner
product on the fibers of the bundle $ad\,X\otimes \Lambda ^{p}(T^{\ast }M),$%
\thinspace $p\geq 1.$

\textit{Case 1:} \ Let $\,p=2$ and let $A$ be a connection 1-form on the
fibers of $X;$ choose $F$ to be the curvature 2-form $F_{A}$ corresponding
to $A.$ \ If $\alpha =0,$ $G=SO(n),$ and $\,n=4,$ then $E_{0}$ is the
Yang-Mills functional.

\textit{Case 2}: \ If $p=2,$ $\alpha =1/2,\,X$ is the bundle of orthonormal
frames on $M,G$ is the Lorentz group $O(1,3),$ and $n=4,$ then $E_{0}$ is
formally analogous to a (torsion-free) gravitational action functional for
the curvature 2-form $F$.

\textit{Case 3}: \ If $p=1,$ $\alpha =1/2,$ $z=A(x,y),$ where $A$ is the
graph of a surface $\Sigma $ in $\mathbf{R}^{3}$, $F=grad\,A$, $X=T^{\ast }M$%
, and $n=2,$ then $E_{1}$ is the energy functional for a nonparametric
minimal surface, as discussed above.

\textit{Case 4}: \ If in the last example $\mathbf{R}^{3}$ is replaced by
the Minkowski space $\mathbf{R}^{n,1}$ and if $p=n-1$, then $E_{1}$ is
closely related to an energy functional for maximal space-like hypersurfaces.%
$^{6}$

With the exception of the last two, these examples are characterized by a
nonvanishing Lie bracket in $A$ due to a nonabelian structure group for $X$.
\ Thus in general these cases are \textit{nonabelian}. \ If $D_{A}$
represents the exterior covariant derivative associated to a connection
1-form $A$ on $ad(X)$ and if $D_{A}^{\ast }$ is the formal adjoint of $D_{A}$%
, then the variational equations for $E_{h}$ can be written$^{8-10}$%
\begin{equation}
D_{A}^{\ast }\left( \rho (Q)F_{A}\right) =0,
\end{equation}
\begin{equation}
D_{A}F_{A}=0.
\end{equation}
The first equation represents a nonabelian version of eq. (4) for curvature
2-forms, and the second equation replaces eq. (5) by the second Bianchi
identity.

An intermediate place between the abelian and nonabelian nonlinear Hodge
theories is occupied by \textit{nonlinear Hodge maps.} \ These are maps $u$
between Riemannian manifolds such that $u$ is a critical point of the
nonlinear Hodge energy (1). \ In this case the geometry of the target space
is enriched in comparison to the abelian case but does not have the
nontrivial Lie group structure of the nonabelian case; the target space is
independent of the base space but is not a curved bundle. \ In the context
of fluid dynamics or the rotation of a nonrigid body, these maps represent
flows on a Riemannian manifold $M$ (typically, a domain of $\mathbf{R}^{n}$)
for which the flow potential is constrained to lie on a possibly different
Riemannian manifold $N$.

Nonlinear Hodge theory can be viewed as an attempt to extend to the
quasilinear field equations of classical physics the unified geometric
treatment given linear field equations by the theory of Hodge and Kodaira. \
The case $\rho =1$ for the abelian equations (4), (5) reduces to the
continuity equation for an incompressible flow $(p=1)$ or the field
equations for electromagnetism $(p=2)$. \ For the nonabelian equations (9),
(10) the case $\rho =1$ reduces to the Yang-Mills equations. \ In the
intermediate case considered in the sequel the case $\rho =1$ reduces to the
equations for harmonic maps (nonlinear sigma-models). In distinction to the
approach in Refs. 1-4, our concern is with the geometry of the target space
rather than the geometry of the domain.

In the following we denote by $C$ generic positive constants, which may
depend on dimension and which may change in value from line to line. \ We
employ the summation convention for repeated indices.

\section{The variational equations}

Consider a map $u:M\rightarrow N$ \ taking a Riemannian manifold $(M,\gamma
) $ into a Riemannian manifold $(N,g).$ \ We are interested in maps which
are critical points of the nonlinear Hodge energy (1). \ Here and throughout
we denote by $x=(x^{1},...,x^{n})$ a coordinate chart on the manifold$\ M$
having metric tensor $\gamma _{\alpha \beta }(x),$ and we denote by $%
u=(u^{1},...,u^{m})$ a coordinate chart on the manifold$\ N$ \ having metric
tensor $g_{ij}(u).$ \ (We assume for the moment that the image of $u$ lies
in a coordinate chart.) The nonlinear Hodge energy assumes the form 
\[
E(u)=\frac{1}{2}\int_{M}\int_{0}^{Q}\rho (s)ds\sqrt{\gamma }dx, 
\]
where 
\[
Q=\gamma ^{\alpha \beta }(x)g_{ij}\left( u(x)\right) \frac{\partial u^{i}}{%
\partial x^{\alpha }}\frac{\partial u^{j}}{\partial x^{\beta }}. 
\]

We have, by the Leibniz rule, 
\begin{equation}
\frac{d}{dt}E(u+t\psi )_{|t=0}=\frac{1}{2}\int_{M}\rho (Q)\frac{d}{dt}%
Q(u+t\psi )_{|t=0}\sqrt{\gamma }dx,
\end{equation}
for arbitrary $\psi \in C_{0}^{\infty }(M).$ \ The construction of the test
functions $\psi $ is not entirely straightforward. \ Use must be made of the
Nash Embedding Theorem to embed $N$ in a higher-dimensional euclidean space.
\ One then employs a nearest point projection $\Pi $ of a suitable euclidean
neighborhood $\mathcal{O}(N)$ onto $N$. \ If $t$ is small enough and $N$ is
a $C^{1}$ submanifold, the variations $\Pi \circ (u+t\psi )$ will be
constrained to lie on $N$, where now $\psi :M\rightarrow \mathcal{O}$. \ A
discussion is given in Section 1 of Ref. 11 for the special case of harmonic
maps.

Carrying out the indicated operation on the right-hand side of (11), we
obtain 
\[
\frac{d}{dt}E(u+t\psi )_{|t=0}=\int_{M}\rho (Q)\gamma ^{\alpha \beta
}(x)g_{ij}\left( u(x)\right) \frac{\partial u^{i}}{\partial x^{\alpha }}%
\frac{\partial \psi ^{j}}{\partial x^{\beta }}\sqrt{\gamma }\,dx 
\]
\begin{equation}
+\frac{1}{2}\int_{M}\rho (Q)\gamma ^{\alpha \beta }(x)\frac{\partial }{%
\partial x^{k}}\left( g_{ij}\left( u(x)\right) \right) \psi ^{k}\frac{%
\partial u^{i}}{\partial x^{\alpha }}\frac{\partial u^{j}}{\partial x^{\beta
}}\sqrt{\gamma }dx.
\end{equation}
But 
\[
\int_{M}\rho (Q)\gamma ^{\alpha \beta }(x)g_{ij}\left( u(x)\right) \frac{%
\partial u^{i}}{\partial x^{\alpha }}\frac{\partial \psi ^{j}}{\partial
x^{\beta }}\sqrt{\gamma }dx= 
\]
\[
\int_{M}\frac{\partial }{\partial x^{\beta }}\left\{ \rho (Q)\sqrt{\gamma }%
\gamma ^{\alpha \beta }(x)g_{ij}\left( u(x)\right) \frac{\partial u^{i}}{%
\partial x^{\alpha }}\psi ^{j}\right\} dx 
\]
\[
-\int_{M}\frac{\partial }{\partial x^{\beta }}\left\{ \rho (Q)\sqrt{\gamma }%
\gamma ^{\alpha \beta }(x)\frac{\partial u^{i}}{\partial x^{\alpha }}%
\right\} g_{ij}\left( u(x)\right) \psi ^{j}dx 
\]
\begin{equation}
-\int_{M}\rho (Q)\gamma ^{\alpha \beta }(x)\frac{\partial }{\partial u^{k}}%
\left( g_{ij}\left( u(x)\right) \right) \psi ^{j}\frac{\partial u^{i}}{%
\partial x^{\alpha }}\frac{\partial u^{k}}{\partial x^{\beta }}\sqrt{\gamma }%
dx.
\end{equation}
Substituting eq. (13) into eq. (12) and taking into account that $\psi $ has
compact support in $M$ \ we obtain, for $\eta ^{i}=g_{ij}\psi ^{j},$ the
formula 
\[
\frac{d}{dt}E(u+t\psi )_{|t=0}=-\int_{M}\frac{\partial }{\partial x^{\beta }}%
\left\{ \rho (Q)\sqrt{\gamma }\gamma ^{\alpha \beta }(x)\frac{\partial u^{i}%
}{\partial x^{\alpha }}\right\} \eta ^{i}dx 
\]
\[
-\frac{1}{2}\int_{M}\rho (Q)\gamma ^{\alpha \beta }(x)g^{\ell j}(u)\left( 
\frac{\partial g_{ij}}{\partial x^{k}}+\frac{\partial g_{kj}}{\partial x^{i}}%
-\frac{\partial g_{ik}}{\partial x^{j}}\right) \frac{\partial u^{i}}{%
\partial x^{\alpha }}\frac{\partial u^{k}}{\partial x^{\beta }}\eta ^{\ell }%
\sqrt{\gamma }dx. 
\]
Applying the definition of affine connection, we conclude that stationary
maps satisfy the system 
\begin{equation}
\frac{1}{\sqrt{\gamma }}\frac{\partial }{\partial x^{\beta }}\left\{ \rho (Q)%
\sqrt{\gamma (x)}\gamma ^{\alpha \beta }(x)\frac{\partial u^{i}}{\partial
x^{\alpha }}\right\} +\rho (Q)\gamma ^{\alpha \beta }(x)\Gamma _{jk}^{i}(u)%
\frac{\partial u^{j}}{\partial x^{\alpha }}\frac{\partial u^{k}}{\partial
x^{\beta }}=0,
\end{equation}
for $i=1,...,m.$

We can also approach the variational equations for nonlinear Hodge maps from
an intrinsic point of view, defining the \textit{nonlinear Hodge tension
field} $\tau $ by the formula 
\[
\tau \equiv trace\,\nabla _{cov}\left( \rho (Q)\omega \right) , 
\]
where $\nabla _{cov}$ denotes the covariant derivative in the bundle $%
T^{\ast }M\otimes u^{-1}TN.$ \ If $N=\mathbf{R}^{k},$ then the equation $%
\tau =0$ reduces to the conventional nonlinear Hodge equation (4) for $%
\omega =du$ [which implies eq. (5)]. \ In particular, if $\Pi \circ (u+t\psi
)$ are the variations described earlier, then we can write the equations for
a weak stationary point in the form 
\begin{equation}
\frac{1}{2}\int_{M}\rho (Q)\left[ \left\langle \nabla u,\nabla \psi
\right\rangle +\nabla _{\psi u}\left\langle D\Pi (u)\left[ \nabla u\right]
,D\Pi (u)\left[ \nabla u\right] \right\rangle _{N}\right] dM=0,
\end{equation}
where 
\begin{equation}
\left| \nabla _{\psi u}\left\langle D\Pi (u)\left[ \nabla u\right] ,D\Pi (u)%
\left[ \nabla u\right] \right\rangle _{N}\right| \leq C\left| \psi \right|
\left| u\right| \left| \nabla u\right| ^{2};
\end{equation}
$\nabla u=grad\;u;$ $\nabla _{\psi u}\left\langle \;,\;\right\rangle _{N}$ \
is the function on $M$ whose value at $x$ is the covariant derivative for $N$
of the metric $\left\langle \;,\;\right\rangle _{N}$ in the direction $\psi
u(x)$. \ (See the discussion leading to inequality (1.2) in Ref. 12; that
paper considers the regularity of energy minimizing maps for $\rho (Q)=Q^{s}$%
.) \ Here we use the fact that for $u\in \Lambda ^{0},$ $\left| \omega
\right| $ is the norm of the gradient of $u$ as well as the norm of its
differential.

The harmonic map density satisfies 
\[
e(u)_{harmonic}\equiv \frac{1}{2}\gamma ^{\alpha \beta }\left\langle \frac{%
\partial u}{\partial x^{\alpha }},\frac{\partial u}{\partial x^{\beta }}%
\right\rangle _{\mid u^{-1}TN}=\frac{1}{2}\left\langle du,du\right\rangle
_{\mid T^{\ast }M\otimes u^{-1}TN}. 
\]
That is, in the harmonic map case the energy density is the trace of the
pullback, via the map $u$, of the metric tensor $g(u)$ on $N$. \ In the case
of nonlinear Hodge maps the situation is a little more complicated, as the
energy density is the integral 
\[
F(\omega )=\int_{0}^{Q}\rho \left( s\right) ds, 
\]
which may not be a quadratic form if $\rho \neq 1.$ \ Moreover, the
nonlinear Hodge density need not scale like a metric tensor. \ Thus the
geometry of harmonic maps is more transparent than the geometry of nonlinear
Hodge maps.

\begin{proposition}
In order for weak solutions $\omega =du$ of the equations $\tau =0$ to exist
locally on $M$ it is sufficient that $N$ be $\mathbf{R}^{k}$ there exist a
positive constant $K<\infty $ for which 
\begin{equation}
K^{-1}\leq \rho (Q)+2Q\rho ^{\prime }(Q)\leq K.
\end{equation}
\end{proposition}

\textit{Proof:} The argument follows Sec. 1 of Ref. 13. \ Define $F(\omega )$
as in the preceding paragraph. \ Then (17) implies that 
\[
\frac{\partial ^{2}F}{\partial \omega _{\beta }\partial \omega _{\alpha }}>0 
\]
and there exist finite positive constants $k_{0}$ and $k_{1}$ such that 
\[
k_{0}Q\leq \frac{\partial ^{2}F}{\partial \omega _{\beta }\partial \omega
_{\alpha }}\omega _{\alpha }\omega _{\beta }\leq k_{1}Q. 
\]
Moreover, there exist finite positive constants $k_{2}$ and $k_{3}$ such
that 
\[
k_{2}Q\leq F\left( Q(\omega )\right) \leq k_{3}Q. 
\]
Thus the energy functional $E$ is convex, bounded above and below, and lower
semicontinuous with respect to weak $L^{2}$ convergence. This completes the
proof of the proposition.

\bigskip

We will use in several contexts the following pointwise inequality for
smooth solutions.

\begin{theorem}
Let $u:M\rightarrow N$ be a $C^{2}$ stationary point of the nonlinear Hodge
energy on $M,$ where $M$ is a compact, $n$-dimensional $C^{\infty }$
Riemannian manifold, $\,n>2,$ and $N$ is a compact $m$-dimensional $%
C^{\infty }$ Riemannian manifold. \ Then the scalar $Q=\left| \nabla
u\right| ^{2}$ satisfies an inequality of the form 
\[
L(Q)+C\left( Q+1\right) Q\geq 0, 
\]
where the second-order operator $L$ is elliptic whenever condition (17) is
satisfied, and the constant $C$ depends on the Ricci curvature of $M$ and
the Riemann curvature of $N$.
\end{theorem}

\textit{Proof.} \ Denote by a subscripted $x^{\sigma }$ differentiation in
the direction of the $\sigma ^{th}$ coordinate. \ Differentiation of the
metric tensor and Christoffel symbols in the direction of an index is
indicated by a comma preceding the subscripted index. \ Choose geodesic
normal coordinates at the points $x\in M$ and $u(x)\in N.$ \ At these points 
\[
\gamma ^{\alpha \beta }(x)=\delta ^{\alpha \beta }(x);\;g_{ij}(u)=\delta
_{ij}(u);\;\Gamma _{\alpha \beta }^{\eta }(x)=\Gamma _{ij}^{k}(u)=0. 
\]
As in the preceding, greek indices are used for coordinates on $M$, and
latin indices, for coordinates on $N$.

Write eq. (14) in the form 
\begin{equation}
\gamma ^{\alpha \beta }\left\{ \rho \left( Q\right) u_{x^{\beta }x^{\alpha
}}^{j}+\rho ^{\prime }\left( Q\right) Q_{x^{\alpha }}u_{x^{\beta }}^{j}-\rho
\left( Q\right) \left[ u_{x^{\eta }}^{j}\Gamma _{\alpha \beta }^{\eta
}(x)-\Gamma _{\ell k}^{j}(u)u_{x^{\alpha }}^{k}u_{x^{\beta }}^{\ell }\right]
\right\} =0.
\end{equation}
Differentiating (18) with respect to $x^{\varepsilon }$ and letting $\alpha
=\beta =\sigma $ yields 
\[
\rho \left( Q\right) u_{x^{\sigma }x^{\sigma }x^{\varepsilon }}^{j}=-\rho
^{\prime }\left( Q\right) Q_{x^{\varepsilon }}u_{x^{\sigma }x^{\sigma
}}^{j}- \left[ \rho ^{\prime }\left( Q\right) Q_{x^{\sigma }}u_{x^{\sigma
}}^{j}\right] _{x^{\varepsilon }} 
\]
\begin{equation}
+\rho \left( Q\right) \left[ u_{x^{\eta }}^{j}\Gamma _{\sigma \sigma
,\varepsilon }^{\eta }(x)-\Gamma _{\ell k,p}^{j}(u)u_{x^{\varepsilon
}}^{p}u_{x^{\sigma }}^{k}u_{x^{\sigma }}^{\ell }\right] .
\end{equation}
Now compute (\textit{c.f.} Sec. 3.2 of Ref. 14) 
\[
\Delta e(u)\equiv \left[ \gamma ^{\alpha \beta }g_{ij}(u)\rho \left(
Q\right) u_{x^{\alpha }}^{i}u_{x^{\beta }}^{j}\right] _{x^{\sigma }x^{\sigma
}}= 
\]
\[
=-\gamma _{\alpha \beta ,\sigma \sigma }g_{ij}(u)\rho \left( Q\right)
u_{x^{\alpha }}^{i}u_{x^{\beta }}^{j}+ 
\]
\[
\gamma ^{\alpha \beta }g_{ij,\ell k}(u)u_{x^{\sigma }}^{\ell }u_{x^{\sigma
}}^{k}\rho \left( Q\right) u_{x^{\alpha }}^{i}u_{x^{\beta }}^{j}+\left[
\gamma ^{\alpha \beta }g_{ij}\rho ^{\prime }\left( Q\right) Q_{x^{\sigma
}}u_{x^{\alpha }}^{i}u_{x^{\beta }}^{j}\right] _{x^{\sigma }}+ 
\]
\[
\left[ \gamma ^{\alpha \beta }g_{ij}\rho \left( Q\right) u_{x^{\alpha
}x^{\sigma }}^{i}u_{x^{\beta }}^{j}\right] _{x^{\sigma }}+\gamma ^{\alpha
\beta }g_{ij}\rho \left( Q\right) u_{x^{\alpha }x^{\sigma }}^{i}u_{x^{\beta
}x^{\sigma }}^{j}+ 
\]
\begin{equation}
\gamma ^{\alpha \beta }g_{ij}\rho \left( Q\right) u_{x^{\alpha
}}^{i}u_{x^{\beta }x^{\sigma }x^{\sigma }}^{j}+\gamma ^{\alpha \beta
}g_{ij}\rho ^{\prime }\left( Q\right) Q_{x^{\sigma }}u_{x^{\alpha
}}^{i}u_{x^{\beta }x^{\sigma }}^{j}\equiv \sum_{s=1}^{7}T_{s}.
\end{equation}
Here 
\begin{equation}
T_{4}=\left[ \gamma ^{\alpha \beta }g_{ij}\rho \left( Q\right) u_{x^{\alpha
}x^{\sigma }}^{i}u_{x^{\beta }}^{j}\right] _{x^{\sigma }}=\left[ \frac{1}{2}%
\rho \left( Q\right) Q_{x^{\sigma }}\right] _{x^{\sigma }}.
\end{equation}
Subtract this term from the left-hand side of eq. (20). \ Applying eq. (19)
with $\varepsilon =\alpha ,$ $\eta =\beta $, we have 
\[
T_{1}+T_{2}+T_{6}+T_{3}= 
\]
\begin{equation}
\rho (Q)\left( R_{\alpha \beta }^{M}u_{x^{\alpha }}^{i}u_{x^{\beta
}}^{i}-R_{ikj\ell }^{N}u_{x^{\alpha }}^{i}u_{x^{\alpha }}^{j}u_{x^{\sigma
}}^{k}u_{x^{\sigma }}^{\ell }\right) +\Lambda (Q),
\end{equation}
where 
\[
\Lambda (Q)\equiv \left\{ \rho ^{\prime }\left( Q\right) u_{x^{\alpha }}^{i} 
\left[ Q_{x^{\sigma }}u_{x^{\alpha }}^{i}-Q_{x^{\alpha }}u_{x^{\sigma }}^{i}%
\right] \right\} _{x^{\sigma }}, 
\]
$R_{\alpha \beta }^{M}$ is the Ricci curvature of $M,$ and $R_{ikj\ell }^{N}$
is the Riemann curvature of $N$. \ The last term in (22) results from
applying the product rule to the quantity $-u_{x^{\alpha }}^{i}\left[ \rho
^{\prime }\left( Q\right) Q_{x^{\sigma }}u_{x^{\sigma }}^{i}\right]
_{x^{\alpha }}.$ \ We can write 
\[
T_{5}=\gamma ^{\alpha \beta }g_{ij}\rho \left( Q\right) u_{x^{\alpha
}x^{\sigma }}^{i}u_{x^{\beta }x^{\sigma }}^{j}=\rho \left( Q\right)
\left\langle \partial _{\sigma }\omega ,\partial _{\sigma }\omega
\right\rangle _{|T^{\ast }M\otimes u^{-1}TN} 
\]
and 
\[
T_{7}=\gamma ^{\alpha \beta }g_{ij}\rho ^{\prime }\left( Q\right)
Q_{x^{\sigma }}u_{x^{\alpha }}^{i}u_{x^{\beta }x^{\sigma }}^{j}=\sum_{\sigma
}2\rho ^{\prime }\left( Q\right) \left\langle \partial _{\sigma }\omega
,\omega \right\rangle _{|T^{\ast }M\otimes u^{-1}TN}^{2}. 
\]
If $\rho ^{\prime }\left( Q\right) \leq 0,$ then (17) and the Schwarz
inequality imply that 
\begin{equation}
T_{5}+T_{7}\geq K^{-1}\left| \nabla \omega \right| ^{2}.
\end{equation}
If $\rho ^{\prime }\left( Q\right) \geq 0,$ then 
\begin{equation}
T_{5}+T_{7}\geq T_{5}=\rho (Q)\left| \nabla \omega \right| ^{2}\geq C\left|
\nabla \omega \right| ^{2}.
\end{equation}
[See the argument leading to (26), below.] \ In either case we obtain from
expressions (20)-(24), defining 
\[
L(Q)\equiv \Delta e(u)-T_{4}-\Lambda (Q) 
\]
\[
=\left\{ \left[ \frac{1}{2}\rho (Q)+Q\rho ^{\prime }\left( Q\right) \right]
Q_{x^{\sigma }}\right\} _{x^{\sigma }}-\Lambda (Q), 
\]
the inequality 
\[
L(Q)\geq \rho (Q)\left( R_{\alpha \beta }^{M}u_{x^{\alpha }}^{i}u_{x^{\beta
}}^{i}-R_{ikj\ell }^{N}u_{x^{\alpha }}^{i}u_{x^{\alpha }}^{j}u_{x^{\sigma
}}^{k}u_{x^{\sigma }}^{\ell }\right) +C\left| \nabla \omega \right| ^{2}. 
\]
\ \ We can write 
\begin{equation}
L(Q)+C\Phi Q\geq 0,
\end{equation}
where $\Phi =Q+1.$

We now show that the operator $L$ is elliptic whenever condition (17) is
satisfied. \ If $\rho ^{\prime }(Q)\leq 0,$ then 
\[
-\rho ^{\prime }\left( Q\right) u_{x^{\alpha }}^{i}\left[ Q_{x^{\sigma
}}u_{x^{\alpha }}^{i}-Q_{x^{\alpha }}u_{x^{\sigma }}^{i}\right] Q_{x^{\sigma
}}= 
\]
\[
\left| \rho ^{\prime }\left( Q\right) \right| u_{x^{\alpha }}^{i}\left[
Q_{x^{\sigma }}u_{x^{\alpha }}^{i}-Q_{x^{\alpha }}u_{x^{\sigma }}^{i}\right]
Q_{x^{\sigma }}= 
\]
\[
\left| \rho ^{\prime }\left( Q\right) \right| \left( Q\left| \nabla Q\right|
^{2}-u_{x^{\alpha }}^{i}u_{x^{\sigma }}^{i}Q_{x^{\alpha }}Q_{x^{\sigma
}}\right) \geq 
\]
\[
\left| \rho ^{\prime }\left( Q\right) \right| \left( Q\left| \nabla Q\right|
^{2}-\left| u_{x^{\alpha }}^{i}u_{x^{\sigma }}^{i}Q_{x^{\alpha
}}Q_{x^{\sigma }}\right| \right) \geq 
\]
\[
\left| \rho ^{\prime }\left( Q\right) \right| \left( Q\left| \nabla Q\right|
^{2}-\frac{1}{2}\left[ \left( u_{x^{\alpha }}^{i}Q_{x^{\alpha }}\right)
^{2}+\left( u_{x^{\sigma }}^{i}Q_{x^{\sigma }}\right) ^{2}\right] \right)
=0, 
\]
where in the final step we have applied Young's inequality. \ \ Thus in this
case 
\[
\left[ \frac{1}{2}\rho (Q)+Q\rho ^{\prime }\left( Q\right) \right] \left|
\nabla Q\right| ^{2}-\rho ^{\prime }\left( Q\right) u_{x^{\alpha }}^{i}\left[
Q_{x^{\sigma }}u_{x^{\alpha }}^{i}-Q_{x^{\alpha }}u_{x^{\sigma }}^{i}\right]
Q_{x^{\sigma }}\geq 
\]
\[
\left[ \frac{1}{2}\rho (Q)+Q\rho ^{\prime }\left( Q\right) \right] \left|
\nabla Q\right| ^{2}\geq \frac{1}{2}K^{-1}\left| \nabla Q\right| ^{2}. 
\]
If $\rho ^{\prime }(Q)\geq 0,$ then $\rho (Q)\geq \rho (Q_{\min })=\rho (0).$
\ At $Q=0$, 
\begin{equation}
\frac{1}{2}\rho (Q)+Q\rho ^{\prime }\left( Q\right) =\frac{1}{2}\rho .
\end{equation}
Condition (17) then implies that $\rho $ is bounded below away from zero. \
Again using Young's inequality, 
\[
\left[ \frac{1}{2}\rho (Q)+Q\rho ^{\prime }\left( Q\right) \right] \left|
\nabla Q\right| ^{2}-\rho ^{\prime }\left( Q\right) u_{x^{\alpha }}^{i}\left[
Q_{x^{\sigma }}u_{x^{\alpha }}^{i}-Q_{x^{\alpha }}u_{x^{\sigma }}^{i}\right]
Q_{x^{\sigma }} 
\]
\[
\geq \left[ \frac{1}{2}\rho (Q)+Q\rho ^{\prime }\left( Q\right) \right]
\left| \nabla Q\right| ^{2}+\rho ^{\prime }\left( Q\right) u_{x^{\alpha
}}^{i}u_{x^{\sigma }}^{i}Q_{x^{\alpha }}Q_{x^{\sigma }}\geq 
\]
\[
\left[ \frac{1}{2}\rho (Q)+Q\rho ^{\prime }\left( Q\right) \right] \left|
\nabla Q\right| ^{2}-Q\rho ^{\prime }\left( Q\right) \left| \nabla Q\right|
^{2}=\frac{1}{2}\rho \left| \nabla Q\right| ^{2}\geq C\left| \nabla Q\right|
^{2}, 
\]
as $\rho $ is bounded below away from zero.

Thus for either sign of $\rho ^{\prime }(Q),$ condition (17) implies that
there is a constant $m_{1}>0$ such that 
\[
\left[ \frac{1}{2}\rho (Q)+Q\rho ^{\prime }\left( Q\right) \right] \left|
\nabla Q\right| ^{2}-\rho ^{\prime }\left( Q\right) u_{x^{\alpha }}^{i}\left[
Q_{x^{\sigma }}u_{x^{\alpha }}^{i}-Q_{x^{\alpha }}u_{x^{\sigma }}^{i}\right]
Q_{x^{\sigma }}\geq m_{1}\left| \nabla Q\right| ^{2} 
\]
and we can write 
\[
L(Q)=\frac{\partial }{\partial x^{j}}\left( a^{ij}(\omega )\frac{\partial Q}{%
\partial x^{i}}\right) 
\]
for a matrix $a^{ij}$ satisfying 
\begin{equation}
m_{1}\left| \xi \right| ^{2}\leq a^{ij}\xi _{i}\xi _{j}\leq m_{2}\left| \xi
\right| ^{2}.
\end{equation}
(See p. 106 of Ref. 3 or Proposition 1.1 of Ref. 4 for a different proof of
ellipticity.) \ This completes the proof of Theorem 2.

\bigskip

The contribution of the target geometry to the nonlinearity of inequality
(25) significantly exceeds that of the geometry of the base manifold, which
in the sequel we generally take to be euclidean.

\section{Apparent singular sets of prescribed dimension}

In general we expect that finite-energy weak solutions to eqs. (15) may be
singular, as singularities occur even in the case $\rho \equiv 1.$ \ It is
natural to seek extra conditions under which solutions are actually smooth.

\begin{theorem}
Let $u:\Omega \rightarrow N$ be a $C^{2}$ stationary point of the nonlinear
Hodge energy on $\Omega /\Sigma ,$ where $\Omega $ is a domain of $\mathbf{R}%
^{n}$, $\,n>2;$ $N$ is a compact $m$-dimensional $C^{\infty }$ Riemannian
manifold, $m\leq n$; $\Sigma \subset \subset B\subset \subset \Omega $ is a
compact singular set, completely contained in a sufficiently small \textit{n}%
-disc $B$, which is itself completely contained in $\Omega .$\ \ Let
conditions (16) and (17) hold. \ If $n>4$, let $2n/(n-2)<\mu \leq n,$ where $%
\mu $ is the codimension of $\Sigma ,$ and let $du$ $\in $ $L^{n}(B).$ \ If $%
n=3,4,$ let $du\in L^{4q_{0}\beta }(B)\cap L^{4q}(B),$ where $\beta =\left(
\mu -\varepsilon \right) /\left( \mu -2-\varepsilon \right) $ for $2<\mu
\leq n,$ $\varepsilon >0,$ and $\frac{1}{2}<q_{0}<q.$ \ Then $du$ is
H\"{o}lder continuous in $\Omega .$
\end{theorem}

\textbf{Remarks.} \ That an $L^{n}$ condition is necessary even for $\rho
\equiv 1$ can be seen by considering the equator map.$^{15}$ \ In Theorem 3
we assume neither that the map $u$ minimizes energy nor that the energy is
small. \ In distinction to the harmonic map case, we do not obtain higher
regularity from the H\"{o}lder continuity of $du$, as the system (15) is not
diagonal in it principal part. In the following proof we assume that $\mu
=n; $ the extension of the proof to lower values of $\mu $ can be effected
by arguments given in Sec. 3 of Ref. 27.

\bigskip

\textit{Proof.} \ Initially let $n$ exceed 4. \ Integrate inequality (25)
against a nonnegative test function $\zeta \in C_{0}^{\infty }(B)$ given by 
\[
\zeta =\left( \eta \psi \right) ^{2}\Xi (Q), 
\]
where $B=B_{R}(x_{0})$ is an $n$-disc, of radius $<R,$ centered at a point $%
x_{0}$ $\in $ $\Omega ;$ assume that $B$ completely contains $\Sigma $ and
is completely contained in $\Omega ;$ $\eta ,\psi \geq 0;\psi (x)=0\,\forall
x$ in a neighborhood of $\Sigma ;\eta \in C_{0}^{\infty }(B^{\prime })$
where $B^{\prime }\subset \subset B$; $\Xi (Q)=H(Q)H^{\prime }(Q),$ where $%
H(Q)=H_{\kappa }(Q)$ is the following variant of Serrin's test function$%
^{16} $: 
\[
H_{\kappa }(Q)= 
\]
\[
\left\{ 
\begin{array}{l}
Q^{[n/(n-2)]^{\kappa }n/4}\;for\;0\leq Q\leq \ell , \\ 
\frac{\mu -\varepsilon }{\mu -2-\varepsilon }\left[ \left( \ell \cdot
Q^{(\mu -2-\varepsilon )/2}\right) ^{[n/(n-2)]^{\kappa }n/2(\mu -\varepsilon
)}-\frac{2}{\mu -\varepsilon }\ell ^{\lbrack n/(n-2)]^{\kappa }n/4}\right]
for\;Q\geq \ell .
\end{array}
\right. 
\]
Iterate the following sequence of elliptic estimates, taking successively $%
u\in L^{\alpha (\kappa )}(B)$ for $\alpha (\kappa )=n[n/(n-2)]^{\kappa
},\kappa =0,1,...\,.$ For all $\kappa <\infty ,$ 
\[
\int_{B^{\prime }}a^{ij}(u)\partial _{i}Q\cdot 2\left( \eta \psi \right)
\partial _{j}\left( \eta \psi \right) \Xi (Q)\ast 1 
\]
\[
+\int_{B^{\prime }}a^{ij}(u)\left( \eta \psi \right) ^{2}\Xi ^{\prime
}(Q)\partial _{i}Q\partial _{j}Q\ast 1\leq \int_{B^{\prime }}\Phi Q\left(
\eta \psi \right) ^{2}\Xi (Q)\ast 1. 
\]
This inequality can be rewritten in the short-hand form 
\begin{equation}
I_{1}+I_{2}\leq I_{3},
\end{equation}
the integrals of which we estimate individually. Because $\mu $ exceeds $%
2n/(n-2)$ we have 
\begin{equation}
\Xi ^{\prime }(Q)\geq C\left( H^{\prime }(Q)\right) ^{2}.
\end{equation}
Also 
\begin{equation}
Q\Xi \leq \left( \frac{n}{n-2}\right) ^{\kappa }\frac{n}{4}H^{2}.
\end{equation}
Inequality (29) implies, by ellipticity, 
\[
I_{2}=\int_{B^{\prime }}a^{ij}(u)\left( \eta \psi \right) ^{2}\Xi ^{\prime
}(Q)\partial _{i}Q\partial _{j}Q\ast 1\geq 
\]
\[
C(m_{1})\int_{B^{\prime }}\left( \eta \psi \right) ^{2}\left( H^{\prime
}(Q)\right) ^{2}\left| \nabla Q\right| ^{2}\ast 1= 
\]
\begin{equation}
C\int_{B^{\prime }}\left( \eta \psi \right) ^{2}\left| \nabla H\right|
^{2}\ast 1\equiv i_{21}.
\end{equation}
Young's inequality implies 
\[
I_{1}=\int_{B^{\prime }}a^{ij}(u)\partial _{i}Q\cdot 2\left( \eta \psi
\right) \partial _{j}\left( \eta \psi \right) H(Q)H^{\prime }(Q)\ast 1= 
\]
\[
2\int_{B^{\prime }}\left( a^{ij}(u)\left( \eta \psi \right) \left( \partial
_{i}H\right) \right) \partial _{j}\left( \eta \psi \right) H\ast 1\geq 
\]
\[
-m_{2}\left( \varepsilon \int_{B^{\prime }}\left( \eta \psi \right)
^{2}\left| \nabla H\right| ^{2}\ast 1+C(\varepsilon )\int_{B^{\prime
}}\left| \nabla \left( \eta \psi \right) \right| ^{2}H^{2}\ast 1\right) 
\]
\begin{equation}
\equiv -\left( i_{11}+i_{12}\right) .
\end{equation}
Using inequality (30) and the Sobolev inequality, we obtain 
\[
I_{3}=\int_{B^{\prime }}\Phi Q\left( \eta \psi \right) ^{2}\Xi (Q)\ast 1\leq
\left( \frac{n}{n-2}\right) ^{\kappa }\frac{n}{4}\int_{B^{\prime }}\Phi
\left( \eta \psi \right) ^{2}H^{2}\ast 1\leq 
\]
\[
C\left\| \Phi \right\| _{n/2}\left( \int_{B^{\prime }}\left( \eta \psi
H\right) ^{2n/(n-2)}\ast 1\right) ^{(n-2)/n}\leq C^{\prime }\left\| \Phi
\right\| _{n/2}\left\| \eta \psi H\right\| _{1,2}^{2} 
\]
\[
\leq C\left\| \Phi \right\| _{n/2}\left\{ \int_{B^{\prime }}\left[ \left|
\nabla \left( \eta \psi \right) \right| ^{2}+\left( \eta \psi \right) ^{2}%
\right] H^{2}\ast 1+\int_{B^{\prime }}\left( \eta \psi \right) ^{2}\left|
\nabla H\right| ^{2}\ast 1\right\} 
\]
\begin{equation}
\equiv i_{31}+i_{32}.
\end{equation}
For sufficiently small $B^{\prime }$we have 
\begin{equation}
0<i_{21}-\left( i_{32}+i_{11}\right) \leq C\left( i_{12}+i_{31}\right) .
\end{equation}
There exists ( \textit{c.f.} Ref. 17, Lemma 2 and p. 73) a sequence of
functions $\xi _{\nu }$ such that:

\textit{a)} $\xi _\nu \in [0,1]\;\forall \nu ;$

\textit{b)} $\xi _{\nu }\equiv 1$ in a neighborhood of $\Sigma ,\;\forall
\nu ;$

\textit{c)} $\xi _{\nu }\rightarrow 0$ \textit{a.e.} as $\nu \rightarrow
\infty ;$

\textit{d)} $\nabla \xi _\nu \rightarrow 0$ in $L^{\mu -\varepsilon }$ as $%
\nu \rightarrow \infty .$

Apply the product rule to the squared $H^{1,2}$ norm in $i_{31}$ letting $%
\psi =\psi _\nu =1-$ $\xi _\nu .$ Observing that the cross terms in $\left(
\nabla \eta \right) \psi $ and $\left( \nabla \psi \right) \eta $ can be
absorbed into the other terms by applying Young's inequality, we estimate 
\[
\lim_{\nu \rightarrow \infty }\int_{B^{\prime }}\eta ^2\left| \nabla \psi
_\nu \right| ^2H^2*1\leq \lim_{\nu \rightarrow \infty }C(\ell
)\int_{B^{\prime }}\left| \nabla \psi _\nu \right| ^2Q^{\frac{\mu
-2-\varepsilon }{\mu -\varepsilon }\left( \frac n{n-2}\right) ^\kappa \frac
n2}*1 
\]
\begin{equation}
\leq \lim_{\nu \rightarrow \infty }C(\ell )\left\| \nabla \psi _\nu \right\|
_{\mu -\varepsilon }^2\left\| u\right\| _{\alpha (\kappa )}^{\alpha (\kappa
)(\mu -2-\varepsilon )/(\mu -\varepsilon )}=0.
\end{equation}
Having shown that the integral on the left in (35) is zero for every value
of $\ell ,$ we can now let $\ell $ tend to infinity. We obtain via Fatou's
Lemma the inequality 
\[
\int_{B^{\prime }}\eta ^2\left| \nabla \left( Q^{\alpha (\kappa )/4}\right)
\right| ^2*1\leq \int_{B^{\prime }}\left| \nabla \eta \right| ^2Q^{\alpha
(\kappa )/2}*1. 
\]
Thus $Q^{\alpha (\kappa )/4}$ is in $H^{1,2}$ on some smaller disc on which $%
\eta =1.$ But then, because $u$ is assumed to be $C^2$ away from the
singularity and $\Sigma $ is compact, $Q^{\alpha (\kappa )/4}$ must be in $%
H^{1,2}$ on the larger disc as well. Apply the Sobolev inequality to
conclude that $u$ is now in the space $L^{\alpha (\kappa +1)}(B).$ Because
the sequence $\left\{ n/(n-2)\right\} ^\kappa $ obviously diverges, we
conclude after a finite number of iterations of this argument that $Q^c$ is
in $H^{1,2}(B)$ for any positive value of $c.$ A final application of the
Sobolev inequality implies that $\omega \in L^s(B)$ for all $s<\infty $ and
for any small $B\subset \subset \Omega ,$ provided $n$ exceeds 4.

Now let $n=3$ or $4$. \ Define$^{16}$%
\[
H_{\kappa }(Q)= 
\]
\[
\left\{ 
\begin{array}{l}
Q^{q\prime }\;for\;0\leq Q\leq \ell , \\ 
\frac{1}{q_{0}}\left[ q^{\prime }\ell ^{q\prime -q_{0}}Q^{q_{0}}-\left(
q_{0}-q^{\prime }\right) \ell ^{q\prime }\right] \;for\;Q\geq \ell ,
\end{array}
\right. 
\]
where $q^{\prime }=\left[ n/\left( n-2\right) \right] ^{\kappa }q.$ \
Arguing as in the higher-dimensional case we obtain, using the Sobolev
inequality, $Q\in L^{2q^{\prime }n/(n-2)}(B).$ \ Repeating the argument for $%
\kappa =0,1,\ldots ,$ we obtain that $\omega \in L^{s}(B)$ for all $s<\infty 
$ when $n$ is 3 or 4.

Now let $n$ be an arbitrary integer greater than 2. \ Again let $\psi _{\nu
}=1-\xi _{\nu },$ where $\xi _{\nu }$ satisfies properties \textit{a)-d)}
above, \ and let $\eta \in C_{0}^{\infty }(B^{\prime })$ as before. \ If $%
\zeta =\eta ^{2}$ $\psi _{\nu },$ then 
\begin{equation}
\int_{B_{R}}\left\langle d\zeta ,\rho (Q)\omega \right\rangle \ast
1=\int_{B_{R}}\left\langle \zeta ,\rho (Q)b(\omega )\right\rangle \ast 1,
\end{equation}
with $b$ given by (15). \ We have 
\[
\int_{B_{R}}\left\langle d\zeta ,\rho (Q)\omega \right\rangle \ast
1=\int_{B_{R}}\left\langle \eta ^{2}\left( d\psi _{\nu }\right) ,\rho
(Q)\omega \right\rangle \ast 1 
\]
\[
+\int_{B_{R}}\left\langle \psi _{\nu }\,d\left( \eta ^{2}\right) ,\rho
(Q)\omega \right\rangle \ast 1, 
\]
where as $\nu $ tends to infinity, $\psi _{\nu }$ tends to $1$ \textit{a.e.}
and 
\[
\left| \int_{B_{R}}\left\langle \eta ^{2}\left( d\psi _{\nu }\right) ,\rho
(Q)\omega \right\rangle \ast 1\right| \leq C(K)\left\| \nabla \psi _{\nu
}\right\| _{\mu -\varepsilon }\left\| \omega \right\| _{\left( \mu
-\varepsilon \right) /\left( \mu -\varepsilon -1\right) }\rightarrow 0. 
\]
Choosing $\eta ^{2}(x)$ to equal 1 for $x\in B_{R/2},$ we find from (36)
that $\omega $ is locally a weak solution in all of $\Omega $.

Let \ the map $\varphi :B_{R}(x_{0})\rightarrow \mathbf{R}^{k}$ satisfy for
sufficiently large $k$ the boundary-value problem 
\begin{eqnarray*}
\delta \left( \rho (Q(d\varphi ))d\varphi \right) &=&0\qquad
in\,\,B_{R}(x_{0}), \\
\varphi _{\vartheta } &=&u_{\vartheta }\qquad on\,\,\partial B,
\end{eqnarray*}
where the subscripted $\vartheta $ denotes the tangential component of the
map in coordinates $(r,\vartheta _{1},...\vartheta _{n-1}).$ \ The existence
of a $C^{2,\alpha }$ solution $\varphi $ to this problem is well known.$^{2}$
\ Moreover, if $(d\varphi )_{R,x_{0}}$ denotes the mean value of the 1-form $%
d\varphi $ on $B_{R}(x_{0}),$ then $d\varphi $ satisfies a \textit{Campanato
estimate}$^{18}$%
\[
\int_{B_{R}(x_{0})}\left| d\varphi -(d\varphi )_{R,x_{0}}\right| ^{2}\ast
1\leq CR^{n+2\gamma _{H}} 
\]
for some number $\gamma _{H}\in (0,1].$ Then $u-\varphi $ is an admissible
test function, and 
\[
\int_{B_{R}(x_{0})}\left\langle d\left( u-\varphi \right) ,\left[ \rho
(Q(du))du-\rho (Q(d\varphi ))d\varphi \right] \right\rangle \ast 1 
\]
\begin{equation}
=\int_{B_{R}(x_{0})}\left\langle \left( u-\varphi \right) ,\rho
(Q)b(u,Du)\right\rangle \ast 1
\end{equation}
with $b$ given by (15). \ Apply to identity (37) Sibner's mean-value formula
(Lemma 1.1 of Ref. 13), which asserts for the unconstrained case that 
\begin{equation}
G^{\alpha }(\xi ,f)-G^{\alpha }(\eta ,h)=A^{\alpha \beta }\left( f_{\beta
}-h_{\beta }\right) +H_{\beta }^{\alpha }\left( \xi ^{\beta }-\eta ^{\beta
}\right) ,
\end{equation}
where 
\[
G^{\alpha }(x,\omega )=\sqrt{\gamma }\frac{\partial F}{\partial \omega
_{\alpha }(x)}, 
\]
$A^{\alpha \beta }$ is a positive-definite matrix, and 
\[
\left| H_{\beta }^{\alpha }\right| \leq C\left( \left| f(x)\right| +\left|
h(x)\right| \right) . 
\]
Here $F$ is the function used in the proof of Proposition 1. \ Equation (38)
extends immediately to our case, as we can estimate the derivative of the
metric $g$ on $N$ by 
\[
\left| \frac{\partial g}{\partial x}\right| \leq \left| \frac{\partial g}{%
\partial u}\right| \left| \frac{\partial u}{\partial x}\right| \leq C\left|
\omega \right| 
\]
[\textit{c.f.} inequality (1.3c) of Ref. 13; the $g$ in the above
inequalities is not the same object as the $g$ in Ref. 13, which corresponds
to our $\gamma $].

In formula (38) choose $\xi =x,\,\eta =0,\,f=\omega ,$ and $h=d\varphi .$ We
obtain, using (16), 
\[
\int_{B_{R}(x_{0})}\left| d\left( u-\varphi \right) \right| ^{2}\ast 1\leq 
\]
\begin{equation}
C\left( \int_{B_{R}(x_{0})}\left( \left| \omega \right| +\left| d\varphi
\right| \right) \left| x\right| \ast 1+\int_{B_{R}(x_{0})}\left| u-\varphi
\right| \rho (Q)\left| u\right| Q\ast 1\right) .
\end{equation}
We can find a number $s$ sufficiently large so that 
\[
\int_{B_{R}(x_{0})}\left( \left| \omega \right| +\left| d\varphi \right|
\right) \left| x\right| \ast 1\leq 
\]
\[
C\left( \left\| \omega \right\| _{s}+\left\| d\varphi \right\| _{s}\right)
\left( \int_{0}^{R}\left| x\right| ^{s/(s-1)}\left| x\right| ^{n-1}d\left|
x\right| \right) ^{(s-1)/s} 
\]
\begin{equation}
\leq C(s,n)R^{[n+s/(s-1)](s-1)/s}\equiv CR^{\eta },
\end{equation}
where\ $\eta >n$ whenever $s>n$. \ Also, Young's inequality yields 
\[
\int_{B_{R}(x_{0})}\left| u-\varphi \right| \rho (Q)\left| u\right| Q\ast
1\leq 
\]
\[
R^{-\nu }\int_{B_{R}(x_{0})}\left| u-\varphi \right| ^{2}\left| u\right|
^{2}\ast 1+R^{\nu }\int_{B_{R}(x_{0})}Q^{2}\rho ^{2}(Q)\ast 1\leq 
\]
\[
R^{-\nu }\int_{B_{R}(x_{0})}\left| u-\varphi \right| ^{2}\left| u\right|
^{2}\ast 1+C\left( \left\| Q\right\| _{s},\left\| \rho \right\| _{\infty
}\right) R^{n(s-1)/s+\nu } 
\]
for a constant $\nu $ to be chosen and $s$ so large that $\nu s>n$. \ We
have 
\[
R^{-\nu }\int_{B_{R}(x_{0})}\left| u-\varphi \right| ^{2}\left| u\right|
^{2}\ast 1\leq 
\]
\[
R^{-\nu }\left( \int_{B_{R}(x_{0})}\left| u-\varphi \right| ^{2n/(n-2)}\ast
1\right) ^{(n-2)/n}\left( \int_{B_{R}(x_{0})}\left| u\right| ^{n}\ast
1\right) ^{2/n} 
\]
\begin{equation}
\leq R^{-\nu }C_{S}\int_{B_{R}(x_{0})}\left| \nabla \left( u-\varphi \right)
\right| ^{2}\ast 1\left( \int_{B_{R}(x_{0})}\left| u\right| ^{n}\ast
1\right) ^{2/n},
\end{equation}
where $C_{S}$ is Sobolev's constant. The $L^{p}$ hypothesis on $du$ now
implies by the Sobolev Theorem (and a trivial application of the
Gaffney-G\aa rding inequality) that for any $\varepsilon >0$ we have 
\[
\int_{B_{R}(x_{0})}\left| u\right| ^{n}\,r^{n-1}drdS\leq 
\]
\[
\left| S^{n}\right| \left( \int_{B_{R}(x_{0})}\left| u\right|
^{n+\varepsilon }\,r^{n-1}dr\right) ^{n/\left( n+\varepsilon \right) }\left(
\int_{B_{R}(x_{0})}\,r^{n-1}dr\right) ^{\varepsilon /\left( n+\varepsilon
\right) }\leq CR^{\lambda } 
\]
for $\lambda =n\varepsilon /(n+\varepsilon )$. \ Because of the high $L^{p}$
space in which $u$ sits we have some flexibility: \ choosing either $%
\varepsilon ,\,s,$ or $\nu $ so that $\nu <\lambda $ allows us to subtract
the right-hand side of inequality (41) from the left-hand side of inequality
(39). \ Because the mean value minimizes variance over all location
parameters, we find that 
\[
\int_{B_{R}(x_{0})}\left| \omega -(\omega )_{R,x_{0}}\right| ^{2}\ast 1\leq
\int_{B_{R}(x_{0})}\left| \omega -(d\varphi )_{R,x_{0}}\right| ^{2}\ast 1 
\]
\[
\leq \int_{B_{R}(x_{0})}\left| \omega -d\varphi \right| ^{2}\ast
1+\int_{B_{R}(x_{0})}\left| d\varphi -(d\varphi )_{R,x_{0}}\right| ^{2}\ast
1 
\]
\[
\leq C\max \left\{ R^{n+2\gamma _{H}},R^{\eta },R^{n(s-1)/s+\nu }\right\} . 
\]
Choosing $x_{0}$ so that $\Sigma \subset \subset B_{R}(x_{0}),$\ and
observing that if $\mu =n,$ this argument works for any smaller positive
value of $R,$ completes the proof.

\section{The sonic limit}

Denote by $\gamma _{1}$ a closed 1-form having prescribed periods. \ We add
to eqs. (4), (5) the \textit{homology condition}$^{1}$ that $\omega -\gamma
_{1}$ be an exact form. \ Denote by $M$ a smooth, compact Riemannian
manifold and consider a family of maps $u_{t}:M\rightarrow N$ into a smooth,
compact Riemannian manifold $N$. \ We further assume that for each $t:0\leq
t<t_{crit},$ $\omega _{t}=du_{t}$ is a \textit{weak minimizer} of the
nonlinear Hodge energy on $M$ in the following sense: \ condition (17) is
satisfied, eqs. (14) are weakly satisfied by the vector field canonically
associated to $\omega _{t}$, $\omega _{t}-t\gamma _{1}$ is an exact form in $%
L^{2}(M),$ and for all other 1-forms $\alpha \in L^{2}(M)$ such that $\alpha
-t\gamma _{1}$ is exact, the inequality 
\[
\int_{M}\int_{0}^{Q(\omega _{t})}\rho (s)ds\,dM\leq
\int_{M}\int_{0}^{Q(\alpha )}\rho (s)ds\,dM 
\]
is satisfied.$^{19}$ \ Borrowing the terminology of fluid dynamics$^{1}$ we
call weak solutions $\omega _{t},\,t\in \lbrack 0,t_{crit}),$ \textit{%
subsonic.} \ The question is whether such solutions converge, as $t$ tends
to $t_{crit},$ to\textit{\ sonic }solutions having velocity $Q_{crit}.$ \
Ellipticity degenerates in the limit as $Q$ tends to $Q_{crit}$ [\textit{c.f.%
} eq. (3) of Section 1]. \ In this limit condition (17) fails and is
replaced by conditions (2), (3). \ In the following theorem we replace $M$
by a euclidean domain; but see the remarks at the end of this section.

\begin{theorem}
Assume the hypotheses of the preceding paragraph. \ That is, let $%
u_{t}:\Omega \rightarrow N$ denote a family of maps between a smooth,
compact domain $\Omega $ of $\mathbf{R}^{n}$ and a coordinate chart on a
smooth, compact m-dimensional Riemannian manifold $N$, $m\leq n,$where $%
0\leq t<t_{crit}$. \ Assume that the 1-forms $\omega _{t}=du_{t}$ weakly
minimize the nonlinear Hodge energy on $\Omega $ over a cohomology class.\
In particular, let the homology condition of the above paragraph be
satisfied for a fixed 1-form $\gamma _{1}.$ \ Assume that the $C^{1}$
function $\rho $ satisfies (2), (3) and that 
\begin{equation}
Q\leq c\int_{\Omega }\int_{0}^{Q}\rho (s)\,ds\;\ast 1\;\;\;\forall Q<Q_{crit}
\end{equation}
for constant $c.$ \ Then as $t$ tends to $t_{crit}$, 
\begin{equation}
\lim_{t\rightarrow t_{crit}}\max_{x\in \,int\,\Omega }Q\left( \omega
_{t}(x)\right) \rightarrow Q_{crit}.
\end{equation}
\end{theorem}

The conclusion of Theorem 4 implies that $\omega _{t}$ depends continuously
on $t$ in the topology of uniform convergence. \ This eventually implies
H\"{o}lder continuity for weak minimizers at the elliptic degeneracy
represented by (3); see Corollary 5.

\bigskip

\textit{Proof.} \ The proof is similar to that of Theorem 4.8 of Ref. 13. \
Denote by $\{t_{\nu }\}$ a nonnegative sequence of points in $[0,t_{crit})$
converging to a limit point. \ We want to establish a sequence of
inequalities satisfied by any subsonic minimizer $\omega _{t_{\nu }}\equiv
\omega _{\nu }.$ Because $\omega _{\nu }$ minimizes energy over a cohomology
class we have 
\[
c\int_{\Omega }\int_{0}^{Q(\omega _{\nu })}\rho (s)ds\,\ast 1 
\]
\[
\leq c\int_{\Omega }\int_{0}^{Q(h_{\nu })}\rho (s)ds\,\ast 1\leq C\left\|
h_{\nu }\right\| _{L^{2}(\Omega )}^{2}, 
\]
where $h_{\nu }$ is a harmonic form such that $h_{\nu }-t_{\nu }\gamma _{1}$
is exact. This gives a uniform bound in $L^{\infty }$ on the sequence $%
\{\omega _{\nu }\}.$\ Now we proceed as in the concluding arguments in the
proof of Theorem 3, comparing $\omega _{\nu }$ to a $C^{1}$ solution $%
\varphi $ of the euclidean nonlinear Hodge equations on $\Omega $. \
Conditions (2) and (42) imply that $u-\varphi $ is an admissible test
function. \ The continuity estimates, starting with formula (37), are also
uniform, as the highest bound imposed on $\omega _{\nu }$\ by these
inequalities is in $L^{n+\varepsilon }.$ \ For example, we can replace
inequality (40) by the estimate 
\[
\int_{B_{R}(x_{0})}\left( \left| \omega \right| +\left| d\varphi \right|
\right) \left| x\right| \ast 1\leq 
\]
\[
C\left( \left\| Q\right\| _{n/2}+\left\| d\varphi \right\| _{n}\right)
\left( \int_{0}^{R}\left| x\right| ^{n/(n-1)}\left| x\right| ^{n-1}d\left|
x\right| \right) ^{(n-1)/n} 
\]
\[
\leq C\left( R^{\delta }+\left( \int_{B_{R}(x_{0})}\left| d\varphi \right|
^{n}\left| x\right| ^{n-1}d\left| x\right| \right) ^{1/n}\right) R^{n} 
\]
\[
\leq C\left( R^{\delta }+\left\| d\varphi \right\| _{ns}\left(
\int_{0}^{R}\left| x\right| ^{n-1}d\left| x\right| \right)
^{(s-1)/ns}\right) R^{n}\leq CR^{\eta } 
\]
for some $\eta >n.$ Thus the hypotheses of Theorem 4 imply the concluding
H\"{o}lder estimate of Theorem 3, from which we obtain equicontinuity for
the sequence $\{\omega _{\nu }\}$. \ Now the Arzel\'{a}-Ascoli Theorem
guarantees uniform convergence of a subsequence to a 1-form satisfying both
the equations and the homology condition. \ This completes the proof of
Theorem 4.

\begin{corollary}
Let the hypotheses of Theorem 4 be satisfied. \ Then $\omega $ is H\"{o}lder
continuous in the interior of $\Omega .$
\end{corollary}

\textit{Proof.} \ Theorem 4 is the crucial ingredient in the technique of 
\textit{Shiffman regularization,} described in the Appendix to Ref. 13. \
This technique is sufficient to establish the H\"{o}lder continuity of $%
\omega $ and prove the corollary.

\bigskip

In Ref. 10 a comparison argument similar to (37)-(41) was constructed for
solutions of eqs. (9), (10). There we chose an exponential gauge at the
origin of coordinates in a euclidean ball $B_{R}(0)$ in order to compare
solutions of (9), (10) with euclidean solutions of (4), (5). \ It was shown
that the difference of the two solutions is small in a high Campanato space.
\ It was then necessary to show that the gauge transformation to an
exponential gauge preserves the Campanato estimate; this allowed us to
extend the comparison outside of $B_{R}(0)$ and apply a covering argument. \
The argument of Ref. 10 provides a guide for extending the results of this
section to maps of a Riemannian manifold $M$. \ The analogy of an
exponential gauge is a choice of geodesic normal coordinates in a local
coordinate chart. \ The arguments of Ref. 13 imply that the difference
between a comparison map $\varphi ,$ taking a euclidean ball into $\mathbf{R}%
^{k}$, and a comparison map $\varphi ^{\prime },$ taking a Riemannian ball
into $\mathbf{R}^{k},$ is itself small in a high Campanato space. \ This is
the analogy of our estimates of the gauge transformations in Ref. 10. \ Now
we can extend the local estimate to all of $M$ by a covering argument. \
Although in principle this method could be used to extend Theorem 3 to a
Riemannian domain as well, in that case no covering argument is needed
because $\Sigma $ is assumed to be small.

\section{An application to harmonic maps}

We now consider the special case in which $\Sigma $ is a point, $\rho $ is
constant, and $n$ exceeds 4. \ The following result is a special case of a
theorem which Liao$^{20}$ proved by quite different methods.

\begin{theorem}
(Liao). \ Let $u:\Omega \rightarrow N$ be a $C^{2}$ stationary point of the
nonlinear Hodge energy with $\rho \equiv 1$ on $\Omega -\{p_{0}\},$ where $%
\Omega $ is a domain of $\mathbf{R}^{n}$, $\,n>4;$ $N$ is a compact $m$%
-dimensional $C^{\infty }$ Riemannian manifold, $m\leq n$; $p_{0}\in \Omega $
is a point. \ If $Q=|du|^{2}$ satisfies the growth condition 
\[
Q(x)\leq \frac{\gamma _{0}}{\left| x-p_{0}\right| ^{2}} 
\]
for $x\in B_{R}(p_{0}),$ where $B_{R}(p_{0})$ is an $n$-disc of radius $R$
centered at $p_{0}$ and $\gamma _{0}$ is a sufficiently small positive
constant, then $du$ is H\"{o}lder continuous on $\Omega .$
\end{theorem}

\textit{Proof.} \ The growth condition guarantees $du\in L^{P}(B)$ $\forall
P<n. $ \ The idea of the proof is to show that $du\in L^{n}(B)$ and apply
Theorem 3. \ Without loss of generality we take $p_{0}$ to lie at the origin
of coordinates in $\mathbf{R}^{n}$.

Let$^{21}$ $\xi (x)=\zeta (x)\psi (x),$ where $x\in B_{R}(0)-\{0\},$%
\[
\psi (x)=|x|^{4-n}, 
\]
and $\zeta $ is chosen so that $\zeta (x)=1$ if $2\varepsilon <|x|\leq R/2,$
and $\zeta (x)=0$ if $|x|<\varepsilon $ or $|x|>R.$ \ We can find $\zeta $
satisfying the additional conditions that 
\[
\left| \nabla \zeta \right| \leq \frac{C}{\varepsilon } 
\]
and 
\[
\left| \Delta \zeta \right| \leq \frac{C}{\varepsilon ^{2}}. 
\]
Because $L$ is a divergence-form operator and $\nabla \xi $ has compact
support in $B_{R}$, inequality (25) implies that 
\[
-\int_{B_{R}(0)}\left( \Delta _{r}\xi \right) Q\ast 1=-\int_{B_{R}(0)}\xi
\left( \Delta Q\right) \ast 1 
\]
\begin{equation}
\leq C\int_{B_{R}(0)}\xi Q^{2}\ast 1,
\end{equation}
where $\Delta _{r}$ is the Laplacian in radial coordinates. \ We have 
\[
\Delta _{r}\xi =\Delta \zeta \cdot \psi +2\nabla \zeta \cdot \nabla \psi
+\zeta \Delta \psi , 
\]
where 
\[
\Delta \psi =2(4-n)\left| x\right| ^{2-n}. 
\]
We can write inequality (44) in the form 
\[
\int_{B_{R}(0)}\psi \zeta Q\left( \frac{-\Delta \psi }{\psi }-CQ\right) \ast
1\leq 2\int_{B_{R}(0)}\left| \nabla \zeta \right| \left| \nabla \psi \right|
Q\ast 1 
\]
\begin{equation}
+\int_{B_{R}(0)}\left| \Delta \zeta \right| \psi Q\ast 1.
\end{equation}
We are interested in the behavior of this inequality as the constant $%
\varepsilon $ in the trapezoidal function tends to zero. \ Write (45) in the
form 
\[
i_{1}\leq 2\,i_{2}+i_{3}. 
\]
Because in $B_{R}(0)-\{0\}$ we have $Q\leq \gamma _{0}\left| x\right| ^{-2},$
integration in radial coordinates yields 
\[
i_{2}\leq \frac{C}{\varepsilon }\int_{\Gamma }d\left| x\right| +C(R), 
\]
where 
\[
\Gamma \equiv \left\{ x|\,\varepsilon \leq \left| x\right| \leq 2\varepsilon
\right\} . 
\]
Integral $i_{2}$ is obviously finite as $\varepsilon $ tends to zero. \
Similarly, \ 
\[
i_{3}\leq \frac{C}{\varepsilon ^{2}}\int_{\Gamma }\left| x\right| d\left|
x\right| +C(R), 
\]
which is also finite for every $\varepsilon .$ \ Finally, 
\[
i_{1}\geq \int_{B_{R}(0)}\left| x\right| ^{4-n}\zeta Q\left( \frac{%
-2(4-n)-C\gamma _{0}}{\left| x\right| ^{2}}\right) \ast 1. 
\]
The quantity inside the largest parentheses on the right is positive
provided $\gamma _{0}$ is sufficiently small. \ In this case 
\[
\lim_{\varepsilon \rightarrow 0}i_{1}\geq C\int_{B_{R/2}(0)}Q\left| x\right|
^{2-n}\ast 1. 
\]
But also, 
\[
\int_{B_{R/2}(0)}Q^{n/2}\ast 1=\int_{B_{R/2}(0)}Q\left( Q^{\left( n-2\right)
/2}\right) \ast 1 
\]
\[
\leq C\int_{B_{R/2}(0)}Q\left| x\right| ^{2-n}\ast 1. 
\]
Taken together, these inequalities imply that $\omega $ lies in the space $%
L^{n}$ in a neighborhood of the singularity. \ The hypotheses of Theorem 3
being satisfied, we conclude that $\omega $ is H\"{o}lder continuous, which
completes the proof of Theorem 6.

\section{Rotational fields}

In this section we study systems of the form 
\begin{equation}
\delta \left( \rho (Q)\omega \right) =0,
\end{equation}
\begin{equation}
d\omega =v\wedge \omega ,
\end{equation}
where $\omega \in \Lambda ^{p}\left( T^{\ast }M\right) $ for $p\geq 1;$%
\textit{\ }$v\in \Lambda ^{1}\left( T^{\ast }M\right) ;$ \textit{M }is an 
\textit{n-}dimensional Riemannian manifold; $Q=\left\langle \omega ,\omega
\right\rangle \equiv \ast (\omega \wedge \ast \omega );$ $\ast :\Lambda
^{p}\rightarrow \Lambda ^{n-p}$ is the Hodge involution; $\rho :\mathbf{R}%
\rightarrow \mathbf{R}^{+}$ is a $C^{1}$ function satisfying the condition$%
^{5}$ 
\begin{equation}
K^{-1}(Q+k)^{q}\leq \rho (Q)+2Q\rho ^{\prime }(Q)\;\leq K(Q+k)^{q}
\end{equation}
for some positive constant $K\;$and nonnegative constants $k,q.$

If $v\equiv 0$ (or if $p=1$ and $v=\omega $), then condition (47)
degenerates to condition (5). \ If $\omega \in \Lambda ^{1}(T^{\ast }M)$ is
the 1-form canonically associated to the velocity field of an \textit{n}%
-dimensional fluid, then condition (5) guarantees that the flow is \textit{%
irrotational:} no circulation exists about any curve homologous to zero.

If $\omega \in \Lambda ^{1}(T^{\ast }M),$ then condition (47) only
guarantees, via the Frobenius Theorem, that $\omega =\ell du$ locally; a
potential exists only along the hypersurfaces $\ell =$ \textit{constant,}
and circulation about topologically trivial points is excluded only along
these hypersurfaces. (For the extension of this result to exterior products
of 1-forms, see, \textit{e.g.,} Ref. 22, Sec. 4-3.) \ Equations (4), (5) can
be used to prescribe a cohomology class for solutions as in Sec. 4, but eqs.
(46), (47) will only prescribe a closed ideal.

We have as an immediate consequence of (47) the condition 
\begin{equation}
d\omega \wedge \omega =0.
\end{equation}
If $\omega $ denotes tangential velocity of a rigid rotor ($\rho =\rho (x)$
only), eq. (49) corresponds in three euclidean dimensions to the fact that
the direction of $\nabla \times \omega $ is perpendicular to the plane of
rotation. Condition (49) also arises in thermodynamics.$^{22,23}$

As in preceding sections, we replace $M$ by a euclidean domain in proving
the technical results. \ In the general case, the curvature of $M$ enters in
a predictable way.$^{24}$

\begin{theorem}
Let $\omega ,v$ smoothly satisfy eqs. (46), (47) on a bounded, open domain $%
\Omega \subset \mathbf{R}^n.$ Assume condition (48). Then the scalar $%
Q=*(\omega \wedge *\omega )$ satisfies the elliptic inequality 
\begin{equation}
L_\omega (Q)+C(Q+k)^q(|\nabla v|+|v|^2)Q\geq 0,
\end{equation}
where $L_\omega $ is a divergence-form operator which is uniformly elliptic
for $k>0.$
\end{theorem}

\textit{Proof.} We have (Ref. 5, (1.5)-(1.7)) 
\[
\left\langle \omega ,\Delta \left( \rho (Q)\omega \right) \right\rangle
=\partial _{i}\left\langle \omega ,\partial _{i}\left( \rho (Q)\omega
\right) \right\rangle -\left\langle \partial _{i}\omega ,\partial _{i}\left(
\rho (Q)\omega \right) \right\rangle 
\]
\begin{equation}
=\Delta H(Q)-\left[ \rho (Q)\left\langle \partial _{i}\omega ,\partial
_{i}\omega \right\rangle +\rho ^{\prime }(Q)\left\langle \partial _{i}\omega
,\omega \right\rangle \partial _{i}Q\right] ,
\end{equation}
where 
\[
\Delta H(Q)=\partial _{i}\left[ \left( \frac{1}{2}\rho (Q)+Q\rho ^{\prime
}(Q)\right) \partial _{i}Q\right] , 
\]
$\partial _{i}=\partial /\partial x^{i},x=x^{1},...,x^{n}\in \Omega $.
Observe that $H$ is defined so that 
\[
H^{\prime }(Q)=\frac{1}{2}\rho (Q)+Q\rho ^{\prime }(Q). 
\]
Just as in the derivation of inequality (25), we have 
\begin{equation}
\rho ^{\prime }(Q)\left\langle \partial _{i}\omega ,\omega \right\rangle
\partial _{i}Q=\sum_{i}2\rho ^{\prime }(Q)\left\langle \partial _{i}\omega
,\omega \right\rangle ^{2}.
\end{equation}
If $\rho ^{\prime }(Q)\geq 0,$ then (52) implies that 
\[
\rho (Q)\left\langle \partial _{i}\omega ,\partial _{i}\omega \right\rangle
+\rho ^{\prime }(Q)\left\langle \partial _{i}\omega ,\omega \right\rangle
\partial _{i}Q\geq 
\]
\begin{equation}
\rho (Q)\left| \nabla \omega \right| ^{2}\geq K^{-1}(Q+k)^{q}\left| \nabla
\omega \right| ^{2}.
\end{equation}
In (53) we have used the inequality 
\begin{equation}
\rho (Q)\geq K^{-1}(Q+k)^{q},
\end{equation}
which follows from (48) (with a possibly larger constant $K)$. If $\rho
^{\prime }(Q)<0,$ then (52) and the Schwarz inequality imply, just as in the
derivation of inequality (25), the inequality 
\[
\rho (Q)\left\langle \partial _{i}\omega ,\partial _{i}\omega \right\rangle
+\rho ^{\prime }(Q)\left\langle \partial _{i}\omega ,\omega \right\rangle
\partial _{i}Q\geq 
\]
\[
\rho (Q)\left| \nabla \omega \right| ^{2}+2\rho ^{\prime }(Q)\left| \nabla
\omega \right| ^{2}Q= 
\]
\begin{equation}
\left[ \rho (Q)+2Q\rho ^{\prime }(Q)\right] \left| \nabla \omega \right|
^{2}\geq K^{-1}(Q+k)^{q}\left| \nabla \omega \right| ^{2}.
\end{equation}
Thus (51) implies, via either (53) or (55) as appropriate, the inequality 
\begin{equation}
\left\langle \omega ,\Delta \left( \rho (Q)\omega \right) \right\rangle \leq
\Delta H(Q)-K^{-1}(Q+k)^{q}\left| \nabla \omega \right| ^{2}.
\end{equation}
Applying eq. (46) to the left-hand side of (56) yields, for $\Delta \equiv
-\left( d\delta +\delta d\right) ,$%
\[
\left\langle \omega ,\Delta \left( \rho (Q)\omega \right) \right\rangle
=-\ast \left[ \omega \wedge \ast \delta d\left( \rho (Q)\omega \right) %
\right] 
\]
\[
=(-1)^{n(p+1)+n}\ast \left[ \omega \wedge \ast (\ast d\ast )d(\rho \omega )%
\right] 
\]
\[
=(-1)^{n\left( n+3\right) -p}\ast \left[ \omega \wedge d\ast d(\rho \omega )%
\right] = 
\]
\[
(-1)^{n\left( n+3\right) }\ast \left\{ d\left[ \omega \wedge \ast d(\rho
\omega )\right] -\left[ d\omega \wedge \ast d(\rho \omega )\right] \right\} 
\]
\begin{equation}
=(-1)^{n(n+3)}\left\{ \ast d\left[ \omega \wedge \ast d(\rho \omega )\right]
-\ast \left[ v\wedge \omega \wedge \ast d(\rho \omega )\right] \right\}
\equiv \tau _{1}-\tau _{2}.
\end{equation}
We express the first term in this difference, up to sign, as a divergence in
the 1-form $dQ,$writing 
\[
\tau _{1}=\ast d\left[ \omega \wedge \ast d(\rho \omega )\right] = 
\]
\begin{equation}
\ast d\left[ \omega \wedge \ast \left( \rho ^{\prime }(Q)dQ\wedge \omega
\right) \right] +\ast d\left[ \omega \wedge \ast \rho d\omega \right] \equiv
\tau _{11}+\tau _{12}.
\end{equation}
Equation (47) implies that 
\[
\tau _{12}\geq -|\tau _{12}|=-\left| \ast d\left[ \omega \wedge \ast (\rho
v\wedge \omega )\right] \right| 
\]
\[
\geq -C\left( |\nabla \omega ||v|\rho |\omega |+|\nabla v|\rho Q+|v||\omega
||\nabla (\rho \omega )|\right) 
\]
\begin{equation}
\equiv C\left( -\tau _{121}-\tau _{122}-\tau _{123}\right) .
\end{equation}
We have, analogously to (54), the inequality $\rho (Q)\leq K(Q+k)^{q}.$
Using this estimate and Young's inequality, we write 
\[
-\tau _{121}=-\sqrt{\rho }|\nabla \omega ||v|\sqrt{\rho }|\omega |\geq 
\]
\begin{equation}
-\varepsilon |\nabla \omega |^{2}(Q+k)^{q}-C(\varepsilon
,K)|v|^{2}(Q+k)^{q}Q.
\end{equation}
Kato's inequality and (48) yield, using $\left| \rho ^{\prime }(Q)\cdot
Q\right| \leq K(Q+k)^{q},$ 
\[
-\tau _{123}=-|v||\omega ||\nabla (\rho \omega )|=-|v||\omega ||\rho
^{\prime }(Q)\nabla Q\cdot \omega +\rho \nabla \omega |\geq 
\]
\[
-|v||\omega |\left( \left| 2\rho ^{\prime }(Q)|\omega |\nabla |\omega |\cdot
\omega \right| +\left| \rho (Q)\nabla \omega \right| \right) \geq 
\]
\[
-2|v||\omega |\left| \rho ^{\prime }(Q)\cdot Q\right| \left| \nabla |\omega
|\right| -|v||\omega |K(Q+k)^{q}\left| \nabla \omega \right| 
\]
\[
\geq -3|v||\omega |K(Q+k)^{q}|\nabla \omega | 
\]
\begin{equation}
\geq -K(Q+k)^{q}\left( \varepsilon |\nabla \omega |^{2}+C(\varepsilon
)|v|^{2}Q\right) .
\end{equation}
Substituting (60) and (61) into (59) yields, for a new $\varepsilon $, 
\[
\tau _{12}\geq -|\tau _{12}|\geq 
\]
\begin{equation}
-K\varepsilon (Q+k)^{q}|\nabla \omega |^{2}-\left( C(\varepsilon
,K)|v|^{2}+K|\nabla v|\right) (Q+k)^{q}Q.
\end{equation}
Similarly, 
\[
\tau _{2}=\ast \left[ v\wedge \omega \wedge \ast d(\rho \omega )\right] \geq
-C|v||\omega ||\nabla (\rho \omega )|, 
\]
which can be estimated by (61). Substituting (62) into (58), (58) into (57),
and (57) into (56), and estimating $\tau _{2}$ of (57) by (61) yields, again
for a new $\varepsilon $, 
\[
\ast d\left[ \omega \wedge \ast \left( \rho ^{\prime }(Q)dQ\wedge \omega
\right) \right] -K\varepsilon (Q+k)^{q}|\nabla \omega |^{2} 
\]
\[
-C(\varepsilon ,K)(Q+k)^{q}\left( |\nabla v|+|v|^{2}\right) Q\leq \Delta
H(Q)-K^{-1}(Q+k)^{q}|\nabla \omega |^{2}. 
\]
We obtain, choosing $0<\varepsilon \leq K^{-2}$ , 
\[
0\leq (K^{-1}-\varepsilon K)(Q+k)^{q}|\nabla \omega |^{2} 
\]
\[
\leq \Delta H(Q)\pm div\left( \ast \left[ \omega \wedge \ast \left( \rho
^{\prime }(Q)dQ\wedge \omega \right) \right] \right) 
\]
\[
+C(Q+k)^{q}\left( |\nabla v|+|v|^{2}\right) Q\equiv L_{\omega
}(Q)+C(Q+k)^{q}\left( |\nabla v|+|v|^{2}\right) Q. 
\]

The ellipticity of the operator $L_{\omega }$ under condition (48) is
obvious from the proof of Theorem 2. This completes the proof of Theorem 7.

\begin{corollary}
Let\thinspace\ $\left( \omega ,v\right) $ be a $C^{2}$ solution of eqs.
(46), (47) on $\Omega /\Sigma ,$ where $\Omega $ is a domain of $\mathbf{R}%
^{n}$, $\,n>2;$ $\Sigma \subset \subset B\subset \subset \Omega $ is a
compact singular set, completely contained in a sufficiently small \textit{n}%
-disc $B$, which is itself completely contained in $\Omega .$\ \ Let
condition (48) hold. \ If $n>4$, let $2n/(n-2)<\mu <n,$ where $\mu $ is the
codimension of $\Sigma ,$ and let $\omega $ $\in $ $L^{n}(B).$ \ If $n=3,4,$
let $\omega \in L^{4\widetilde{q}_{0}\beta }(B)\cap L^{4\widetilde{q}}(B),$
where $\beta =\left( \mu -\varepsilon \right) /\left( \mu -2-\varepsilon
\right) $ for $2<\mu \leq n,$ $\varepsilon >0,$ and $\frac{1}{2}<\widetilde{q%
}_{0}<\widetilde{q}.$ \ If \ $(Q+k)^{q}\left( |\nabla v|+|v|^{2}\right) \in
L^{n/2}(B)$ and $|\nabla v|+|v|^{2}\in L^{p}(B)$ for some $p$ exceeding $n/2$%
, then $\omega $ is bounded on compact subdomains of $\Omega .$ This bound
is uniform for $k>0.$\ 
\end{corollary}

\textit{Proof.} \ In (25) take $\Phi =\left( Q+k\right) ^{q}\left( \left|
\nabla v\right| +\left| v\right| ^{2}\right) .$ \ Apply the arguments
leading to (36) to show that $Q$ is an $H^{1,2}$ weak solution. \ Now choose$%
^{13}$%
\[
\zeta =\left( \left| \omega _{k}\right| +\delta \right) ^{2\tau -2}\eta ^{2} 
\]
for $\left\{ \omega _{k}\right\} $ an increasing sequence chosen so that $%
\lim_{k\rightarrow \infty }\omega _{k}=\omega ;$ $\eta \in C_{0}^{\infty
}(B);$ $\eta \geq 0;$ $\delta >0;$ $\tau >1.$ Estimating (36) for this
choice of test function implies in the limit that $\left| \omega \right|
^{\tau }\in H^{1,2}(B)$ for some $\tau >1.$ \ Also, $\left( \left| \omega
\right| ^{\tau }\right) ^{\lambda }$ satisfies (25) for $\lambda <2.$ \ Now
Theorem 5.3.1 of Ref. 19 implies that $\left| \omega \right| $ is bounded.

\section{The heat flow of solutions}

Consider the system 
\begin{equation}
-\delta \left( \rho \left( Q(x,t)\right) \omega (x,t)\right) =\frac{\partial
u(x,t)}{\partial t},
\end{equation}
\begin{equation}
du(x,t)=\omega (x,t),
\end{equation}
where $x\in M,\,t\in (0,T],$ and exterior differentiation is in the space
directions only. \ Solutions of eqs. (63), (64) describe the \textit{heat
flow,} or \textit{gradient flow,} of nonlinear Hodge maps. \ Notice that
(64) implies $d\omega =0.$

If $M$ is compact or if the normal component of $\omega $ vanishes on $%
\partial M,$ then the time decay of the energy 
\[
E_{t}(\omega )\equiv \frac{1}{2}\int_{M}\int_{0}^{Q\left( \omega
(x,t)\right) }\rho (s)ds\,dM 
\]
is given by 
\[
\frac{d}{dt}E_{t}(\omega )=\frac{1}{2}\int_{M}\rho (Q)\frac{\partial Q}{%
\partial t}\,dM= 
\]
\[
\int_{M}\rho (Q)\left\langle \frac{\partial \omega }{\partial t},\omega
\right\rangle \,dM=\int_{M}\left\langle \frac{\partial \omega }{\partial t}%
,\rho (Q)\omega \right\rangle \,dM. 
\]
Equations (63), (64) imply that 
\begin{equation}
\frac{\partial \omega }{\partial t}=\frac{\partial (du)}{\partial t}=d\left( 
\frac{\partial u}{\partial t}\right) =-d\delta \left( \rho (Q)\omega \right)
.
\end{equation}
These identities together imply that 
\[
\frac{d}{dt}E_{t}(\omega )=-\int_{M}\left\langle d\delta \left( \rho
(Q)\omega \right) ,\rho (Q)\omega \right\rangle \;dM 
\]
\begin{equation}
=-\int_{M}\left\langle \delta \left( \rho (Q)\omega \right) ,\delta \left(
\rho (Q)\omega \right) \right\rangle \;dM\leq 0.
\end{equation}
We conclude from (66) that a finite energy functional will remain so
indefinitely.

The local estimate for $Q,$ taking $M$ to be a bounded, open domain of $%
\mathbf{R}^{n}$, is similar to its elliptic counterparts in the proof of
Theorem 7: If $\rho \left( Q(x,t)\right) $ satisfies inequality (48), then 
\[
\partial _{i}\left[ \left( \frac{1}{2}\rho (Q)+Q\rho ^{\prime }(Q)\right)
\partial _{i}Q\right] -K^{-1}(Q+k)^{q}\left| \nabla \omega \right| ^{2}\geq 
\]
\[
\left\langle \omega ,\Delta \left( \rho \omega \right) \right\rangle =\ast d 
\left[ \omega \wedge \ast \left( \rho ^{\prime }(Q)dQ\wedge \omega \right) %
\right] +\ast \left[ \omega \wedge \ast \frac{\partial \omega }{\partial t}%
\right] 
\]
using (63), (64), and (65), and 
\[
0\leq K^{-1}(Q+\kappa )^{q}\left| \nabla \omega \right| ^{2}\leq L_{\omega
}(Q)-\frac{1}{2}\frac{\partial Q}{\partial t}\equiv 
\]
\begin{equation}
\partial _{i}\left[ \left( \frac{1}{2}\rho (Q)+Q\rho ^{\prime }(Q)\right)
\partial _{i}Q\right] \pm div\ast \left[ \omega \wedge \ast \left( \rho
^{\prime }(Q)dQ\wedge \omega \right) \right] -\frac{1}{2}\frac{\partial Q}{%
\partial t}.
\end{equation}
This inequality is uniformly subparabolic whenever condition (48) is
satisfied for $k>0$ or $q=0.$

If $M$ is a compact Riemannian manifold and $u:M\times \lbrack
0,T]\rightarrow N,$ then we can obtain a global estimate for $Q$. \ In place
of (14) we have the parabolic system 
\[
\frac{1}{\sqrt{\gamma }}\frac{\partial }{\partial x^{\beta }}\left\{ \rho (Q)%
\sqrt{\gamma }\gamma ^{\alpha \beta }\frac{\partial u^{i}}{\partial
x^{\alpha }}\right\} +\rho (Q)\gamma ^{\alpha \beta }\Gamma _{jk}^{i}(u)%
\frac{\partial u^{j}}{\partial x^{\alpha }}\frac{\partial u^{k}}{\partial
x^{\beta }}=u_{t}^{i}. 
\]
Arguing as in the proof of Theorem 2, we add to the middle and right-hand
side of eq. (20) a term of the form 
\[
u_{x^{\alpha }}^{i}u_{tx^{\alpha }}^{i}=u_{x^{\alpha }}^{i}u_{x^{\alpha
}t}^{i}=\frac{1}{2}Q_{t}. 
\]
Let the sectional curvature of $N$ be nonpositive. We obtain as in (67) the
inequality 
\begin{equation}
L_{\omega }(Q)+C_{R}\rho Q-\frac{1}{2}\frac{\partial Q}{\partial t}\geq 0,
\end{equation}
where $C_{R}$ depends on the Ricci curvature of $M.$\ This inequality is, of
course, also uniformly subparabolic whenever condition (17) is satisfied. In
fact we have an \textit{a priori} estimate in this case, which strongly
depends on the ellipticity of $L_{\omega }$.

\begin{theorem}
Let $u(x,t)$ be a mapping of a smooth, compact $n$-dimensional Riemannian
cylinder $M\times \lbrack 0,T]$ into $N,$ where $N$ is a smooth, compact
Riemannian manifold of nonpositive sectional curvature and $T$ is a finite
number. Suppose that $u$ smoothly satisfies (68) with $\rho $ small in $%
L^{s/2}(M),$ $s>n,$ and with $\rho ^{\prime }(\widetilde{s})\leq 0,$ $%
\widetilde{s}\in \lbrack 0,Q]$. \ Let condition (17) hold for each point $%
\left( x,t\right) \in M\times \lbrack 0,T]$ and let $Q(x,0)\leq 1.$ Let the $%
H^{1,2}$ Sobolev inequality hold on $M$ for constants $S_{1}$, $S_{2}$. \
Then there is a constant $c(s,K,q,M,N,T,S_{1},S_{2})$ such that for $q>0,$ 
\[
\sup_{t\in (0,T]}\left( \sup_{x\in M}Q(x,t)\right) \leq ct^{-n/2(q+1)}\left(
E\left[ \omega (x,0)\right] \right) ^{1/(q+1)}, 
\]
where $E$ is the nonlinear Hodge energy.
\end{theorem}

\textit{Proof.} \ Multiply inequality (68) by $\left( Q+\beta \right) ^{r-1}$
for $r>1$ and $\beta >0.$ Replace the time derivative in (68) by the
(identical) time derivative of $Q+\beta $ and integrate over $M$. \ We
obtain 
\[
r^{-1}\frac{\partial }{\partial t}\int_{M}\left( Q+\beta \right)
^{r}\,dM\leq \int_{M}\left( Q+\beta \right) ^{r-1}\nabla \cdot \left(
a\left( \omega \right) \nabla Q\right) \,dM 
\]
\begin{equation}
+C\int_{M}\rho \cdot \left( Q+\beta \right) ^{r}\,dM,
\end{equation}
where $\nabla $ is the gradient on $M$ and $a$ is the matrix-valued function
of inequality (27). \ Because $M$ is compact, Stokes' Theorem implies that 
\[
\int_{M}\left( Q+\beta \right) ^{r-1}\nabla \cdot \left( a\left( \omega
\right) \nabla Q\right) \,dM=\int_{M}\nabla \cdot \left( a\left( \omega
\right) \left( Q+\beta \right) ^{r-1}\nabla Q\right) \,dM 
\]
\[
-\int_{M}\nabla \left( \left( Q+\beta \right) ^{r-1}\right) \cdot a\left(
\omega \right) \nabla Q\,dM=-\int_{M}\left( r-1\right) \left( Q+\beta
\right) ^{r-2}a\left( \omega \right) \left| \nabla Q\right| ^{2}\,dM 
\]
\[
\leq -m_{1}\int_{M}\left( r-1\right) \left( Q+\beta \right) ^{r-2}\left|
\nabla Q\right| ^{2}\,dM=-m_{1}\int_{M}\nabla \left( \left( Q+\beta \right)
^{r-1}\right) \cdot \nabla Q\,dM, 
\]
where $m_{1}$ depends on $k,K,$ and $q$. \ Now 
\[
-\left| \nabla \left( \left( Q+\beta \right) ^{r/2}\right) \right|
^{2}=-\left| \frac{r}{2}\left( Q+\beta \right) ^{(r-2)/2}\nabla Q\right|
^{2}= 
\]
\[
-\frac{r^{2}}{4}\left( Q+\beta \right) ^{r-2}\left( \nabla Q\right) ^{2}=-%
\frac{r^{2}}{4\left( r-1\right) }\nabla \left( \left( Q+\beta \right)
^{r-1}\right) \nabla Q, 
\]
so we can write inequality (69) in the form 
\[
r^{-1}\frac{\partial }{\partial t}\int_{M}\left( Q+\beta \right)
^{r}\,dM\leq -\frac{4m_{1}(r-1)}{r^{2}}\int_{M}\left| \nabla \left( \left(
Q+\beta \right) ^{r/2}\right) \right| ^{2}dM 
\]
\[
+C\left\| \rho \right\| _{s/2}\left\| \left( Q+\beta \right) ^{r}\right\|
_{s/(s-2)}^{2}. 
\]
Employing the parabolic DeGiorgi-Nash-Moser iteration as in Sec. 4 of Ref.
25, taking $p_{0}=q+1$, we obtain, letting $\beta $ tend to zero, 
\[
\sup_{t\in (0,T]}\left( \sup_{x\in M}Q(x,t)\right) \leq Ct^{-n/2(q+1)}\left(
\int_{M}\left| Q(x,0)\right| ^{q+1}dM\right) ^{1/(q+1)}. 
\]
Because 
\[
\frac{d}{ds}\left( s\rho (s)\right) =\rho (s)+s\rho ^{\prime }(s)\leq \rho
(s), 
\]
we have, for $\rho ^{\prime }(s)\leq 0,$ the inequality 
\[
Q\rho (Q)=\int_{0}^{Q}\frac{d}{ds}\left( s\rho (s)\right) ds\leq
\int_{0}^{Q}\rho (s)\,ds. 
\]
Thus 
\[
2E_{|t=0}\geq \int_{M}\int_{0}^{Q(x,0)}\rho (s)ds\,dM\geq \int_{M}Q(x,0)\rho
(Q,0)dM\geq 
\]
\[
\int_{M}Q(x,0)\left[ \rho (Q,0)+2Q(x,0)\rho ^{\prime }(Q)\right] dM\geq
K^{-1}\int_{M}Q(x,0)^{q+1}dM. 
\]
Taking the $\left( q+1\right) ^{st}$ root of this inequality and using (66)
completes the proof of Theorem 9.

\bigskip

A local version of Theorem 9 would argue from inequality (67) rather than
(68). \ The initial argument is as in the proof of Theorem 9 except that the
integration is against cut-off functions. \ The Moser iteration is
implemented as in Ref. 26.

\bigskip \ 

$^{1}$L. M. Sibner and R. J. Sibner, A nonlinear Hodge-de Rham theorem, 
\textit{Acta Math.} \textbf{125}, 57-73 (1970).

$^{2}$L. M. Sibner and R. J. Sibner, Nonlinear Hodge theory: Applications, 
\textit{Advances in Math.} \textbf{31}, 1-15 (1979).

$^{3}$L. M. Sibner and R. J. Sibner, A maximum principle for compressible
flow on a surface, \textit{Proc. Amer. Math. Soc.} \textbf{71}(1), 103-108
(1978).

$^{4}$L. M. Sibner and R. J. Sibner, A subelliptic estimate for a class of
invariantly defined elliptic systems, \textit{Pacific J. Math.} \textbf{94}%
(2), 417-421 (1981).

$^{5}$K. Uhlenbeck, Regularity for a class of nonlinear elliptic systems, 
\textit{Acta Math.} \textbf{138}, 219-240\textbf{\ }(1977).

$^{6}$Y. Yang, Classical solutions in the Born-Infeld theory, \textit{Proc.
R. Soc. Lond. Ser. A} \textbf{456}, no. 1995, 615-640 (2000).

$^{7}$L. Bers, \textit{Mathematical Aspects of Subsonic and Transonic Gas
Dynamics} (Wiley, New York, 1958).

$^{8}$T. H. Otway, Yang-Mills fields with nonquadratic energy, \textit{J.
Geometry \& Physics} \textbf{19}, 379-398\textbf{\ }(1996).

$^{9}$T. H. Otway, Properties of nonlinear Hodge fields, \textit{J. Geometry
\& Physics }\textbf{27}, 65-78 (1998).

$^{10}$T. H. Otway, Nonlinear Hodge structures in vector bundles, in: 
\textit{Nonlinear Analysis in Geometry and Topology,} edited by Th. M.
Rassias (Hadronic Press, Palm Harbor, 2000); see also arXiv:math-ph/9808013.

$^{11}$R. Schoen, Analytic aspects of the harmonic map problem, in: \ 
\textit{Seminar in Nonlinear Partial Differential Equations,} edited by S.
S. Chern (Springer, New York, 1984).

$^{12}$R. Hardt and F-H. Lin, Mappings minimizing the $L^{p}$ norm of the
gradient, \textit{Commun. Pure Appl. Math.} \textbf{40,} 555-588 (1987).

$^{13}$L. M. Sibner, An existence theorem for a nonregular variational
problem, \textit{Manuscripta} \textit{Math.} \textbf{43}, 45-72\textbf{\ }%
(1983).

$^{14}$J. Jost, \textit{Nonlinear Methods in Riemannian and K\"{a}hlerian
Geometry} (Birkh\"{a}user, Basel, 1988).

$^{15}$S. Takakuwa, On removable singularities of stationary harmonic maps, 
\textit{J. Fac. Sci. Univ. Tokyo,} Sect. IA, Math. \textbf{32}, 373-395
(1985); \ S. Hildebrandt, in \ \textit{Proceedings of the 1980 Beijing
Symposium on Differential Geometry and Differential Equations}, edited by S.
S. Chern and Wu Wen-ts\"{u}n, Science Press, Beijing, and Gordon and Breach,
New York, 1982.

$^{16}$J. Serrin, Local behavior of solutions of quasilinear equations, 
\textit{Acta Math.} \textbf{111}, 247-302 (1964).

$^{17}$J. Serrin, Removable singularities of solutions of elliptic
equations, \textit{Archs. Ration. Mech. Analysis} \textbf{17}, 67-78 (1964).

$^{18}$M. Giaquinta, \textit{Multiple Integrals in the Calculus of
Variations and Nonlinear Elliptic Systems} (Princeton University Press,
Princeton, 1983).

$^{19}$C. B. Morrey, \textit{Multiple Integrals in the Calculus of Variations%
} (Springer, Berlin, 1966).

$^{20}$G. Liao, A regularity theorem for harmonic maps with small energy, 
\textit{J. Differential Geometry} \textbf{22}, 233-241 (1985).

$^{21}$B. Gidas and J. Spruck, Global and local behavior of positive
solutions of nonlinear elliptic equations, \textit{Commun. Pure Appl. Math.} 
\textbf{34,} 525-598 (1981); T. H. Otway and L. M. Sibner, Point
singularities of coupled gauge fields with low energy, \textit{Commun. Math.
Physics} \textbf{111}, 275-279 (1987).

$^{22}$D. G. B. Edelen, \textit{Applied Exterior Calculus} (Wiley, New York,
1985).

$^{23}$C. Carath\'{e}odory, \textit{Gesammelte Mathematische Schriften}, Bd.
II, S. 131-177 (C. H. Beck'sche Verlagsbuchhandlung, M\"{u}nich, 1955).

$^{24}$A. Lichnerowicz, Courbure et nombres de Betti d'une variet\'{e}
riemannienne compacte, \textit{C. R. Acad. Sci. Paris} \textbf{226,}
1678-1680 (1948).

$^{25}$D. Yang, $L^{p}$ pinching and compactness theorems for compact
Riemannian manifolds, preprint.

$^{26}$D. Yang, Convergence of Riemannian manifolds with integral bounds on
curvature I, \textit{Ann. scient. \'{E}c. Norm. Sup.} \textbf{25}, 77-105
(1992).

$^{27}$T. H. Otway, Maps and fields with compressible density, \textit{%
Rediconti Sem. Mat. Universit\`{a} Padova} \textbf{111}, (2004), to appear.

\end{document}